\documentclass{aa}  

\usepackage{graphicx}
\usepackage{txfonts,textcomp}
\usepackage{style_paper}
\usepackage{hyperref}
\usepackage{gensymb}
\usepackage[flushleft]{threeparttable}
\usepackage{multirow}
\begin{document}

%%%%%%%%%%%%%%%%%%%%%%%%%%%%%%%%%%%%%%%%

    \title{The circumbinary disk of HD\,34700A}
    \subtitle{I. CO gas kinematics indicate spirals, \\ infall, and vortex motions}
   %\thanks{The ALMA 12CO molecular line emission cube is only available in electronic form at the CDS via anonymous ftp to \url{cdsarc.cds.unistra.fr} (130.79.128.5) or via \url{https://cdsarc.cds.unistra.fr/cgi-bin/qcat?J/A+A/}}}
   \author{J. Stadler\inst{1,2}, M. Benisty\inst{3}, F. Zagaria\inst{3}, A. F. Izquierdo\inst{4,5}, J. Speedie\inst{6,7,8}, A. J. Winter\inst{9,3}, L. W{\"o}lfer\inst{7}, \\
   J. Bae\inst{4}, S. Facchini\inst{10},  D. Fasano\inst{1,3}, N. Kurtovic\inst{11}, R. Teague\inst{7}}

   \institute{
   $^{1}$ Universit\'{e} C\^{o}te d'Azur, Observatoire de la C\^{o}te d'Azur, CNRS, Laboratoire Lagrange, France\\
   $^{2}$ European Southern Observatory, Karl-Schwarzschild-Str. 2, 85748 Garching bei München, Germany\\
   \email{jochen.stadler@eso.org} \\
   $^{3}$ Max-Planck Institute for Astronomy (MPIA), Königstuhl 17, 69117 Heidelberg, Germany \\
   $^{4}$ Department of Astronomy, University of Florida, Gainesville, FL 32611, USA \\
   $^{5}$ NASA Hubble Fellowship Program Sagan Fellow \\
   $^{6}$ Department of Physics \& Astronomy, University of Victoria, Victoria, BC, V8P 5C2, Canada \\
   $^{7}$ Department of Earth, Atmospheric, and Planetary Sciences, Massachusetts Institute of Technology, Cambridge, MA 02139, USA\\ 
    $^{8}$ Heising-Simons Foundation 51 Pegasi b Fellow \\
   $^{9}$ Astronomy Unit, School of Physics and Astronomy, Queen Mary University of London, London E1 4NS, UK \\
   $^{10}$ Universit\'a degli Studi di Milano, via Celoria 16, 20133 Milano, Italy \\
   $^{11}$ Max Planck Institute for Extraterrestrial Physics, Giessenbachstrasse 1, D-85748 Garching, Germany \\
   }
   
   \date{Submitted on July 29 2025; accepted for publication in A\&A on January 16, 2026.}

% \abstract{}{}{}{}{}
% 5 {} token are mandatory

  \abstract
   {}%The formation of planets in multi-star systems is significantly influenced by the interactions between the protoplanetary disks and their stellar hosts, as well as the surrounding environment. 
   %Despite the discovery of hundreds of planets in these systems, we still need to comprehend the mechanisms that influence the ability of such disks to form planets.}
   {We present the first high-resolution ($\sim0\farcs14$) Atacama Large Millimeter/submillimeter Array (ALMA) Band 6 dust continuum, \twCOfull{}, \thCOfull{}, and \eiCOfull{} molecular line emission observations of the quadruple system HD\,34700. In particular, HD\,34700AaAb is a spectroscopic binary ($M_{\rm{bin}}=4\,M_\odot$) surrounded by two low-mass companions (B$=0.6\,M_\odot$, C$=0.4\,M_\odot$) at large separations. Its circumbinary disk is highly substructured, featuring numerous spiral arms and a large cavity observed in infrared (IR) scattered light. We aim to shed light on the nature of these features by examining the gas kinematics at work in the circumbinary disk.}
   {We analyzed the CO line channel and intensity moment maps. By fitting a Keplerian model to the line channel emission, we identified the residual motions and conducted a line spectra analysis.}
   % Results
   {We resolved an asymmetric continuum crescent on top of a dust ring at $0\farcs39$ (138\,au) colocated with the IR ring. The CO molecule's line emission traces a smaller cavity in gas, whose edge aligns with the inner rim of the ring detected in H$\alpha$ emission at $0\farcs20$ (65\,au). The \twCO{} line emission and kinematics trace highly non-Keplerian motions ($\sim0.1\Delta\upsilon_k$) and these CO spiral features align well with the spiral structures in scattered light. The \twCO{} line spectra analysis reveals a streamer above the southeastern disk plane, likely falling onto the disk. The \thCO{} and \eiCO{} kinematics largely follow the disk's underlying Keplerian rotation, while \thCO{} exhibits tentative signs of anticyclonic vortex flows at the continuum crescent location.}
   % Conclusions
   {Our multimolecular line study suggests that the circumbinary disk of HD\,34700A is highly perturbed in its upper layers, possibly warped and influenced by infalling material. While late-stage infall may account for the IR spirals and the formation of the vortex through Rossby wave instability, an embedded massive companion within the cavity might also be contributing to these features.}


   \titlerunning{The complex kinematics of HD\,34700A}
   \authorrunning{Stadler et al.}
   \maketitle
%
%-------------------------------------------------------------------
%%%%%%%%%%%%%%%%%%%%%%%%%%%%%%%%%%%%%%%%%%%%%%
 \section{Introduction}
%%%%%%%%%%%%%%%%%%%%%%%%%%%%%%%%%%%%%%%%%%%%%%
A high fraction of stars are born in multiple systems \citep{Offner_2023_PPVII}. The presence of stellar companions can significantly influence the evolution of protoplanetary disks. Tidal interactions can affect the morphology of protoplanetary disks through the formation of spirals, misalignment, and truncation \citep[see][for a recent review]{cuello2025}. Consequently, these gravitational interactions are crucial in determining the disks' potential for planetary formation \citep{Marazi_Thebault_2019}. Planets have been observed in multiple systems, with over 700 planets detected around single stars (S-type) and over 30 orbiting a binary star \citep[P-type,][]{Thebault_Bonanni_2025}. In circumbinary disks (CBDs), the tidal forces created by the binary system carve out a cavity in the inner region of the disk, whose size is roughly a few times the separation of the binary \citep{Artymowicz_ea_1994, Hirsch_ea_2020, Penzlin_ea_2024}. In the case of a high mass ratio in binary systems ($q = M_2/M_1 > 0.05$), the cavity becomes eccentric and develops horseshoe-shaped structures \citep[e.g.,][]{Ragusa_ea_2020}. Identifying such substructures in gas and dust observations can provide clues about the presence of massive companions \citep[e.g.,][]{Ragusa_ea_2021, Kurtovic_ea_2022, Calcino_ea_2023, Calcino_ea_2024}.

\begin{figure*}[t!]
    \centering
    \includegraphics[width=1\linewidth]{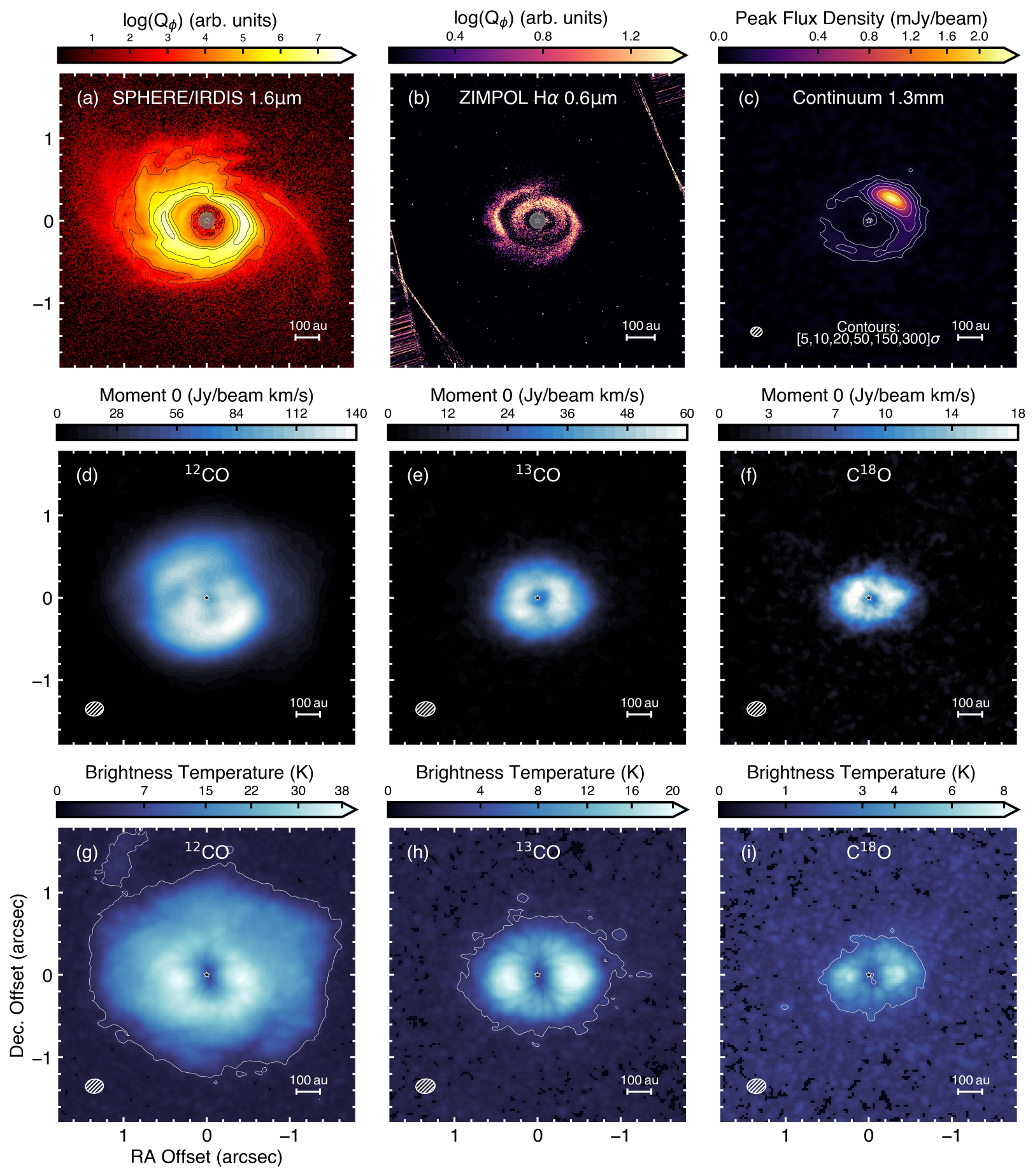}
    \caption{Multi-wavelengths observations of \sys{}. \textbf{(a)} SPHERE/IRDIS IR scattered light observations and \textbf{(b)} ZIMPOL H$\alpha$ observations presented in \citet{Columba_ea_2024}, with the gray circles representing the coronograph ($r=92.5\,$mas\,$\approx32\,$au). \textbf{(c)} ALMA 225.3~GHz continuum observations. \textbf{(d)}, \textbf{(e),} and \textbf{(f)} show the moment 0 maps for the \twCO, \thCO,  and \eiCO fiducial cubes, while \textbf{(g)}, \textbf{(h),} and \textbf{(i)} show the peak
intensity maps for the same cubes, respectively. The brightness temperature was computed using the Rayleigh-Jeans approximation. The white contour in the peak intensity maps encloses 5$\sigma$ of the root-mean-square (RMS) noise of each line. The location of the spectroscopic binary is highlighted with the star. The beam size is shown in the lower left corner for the ALMA observations.}
    \label{fig:overview}
\end{figure*}

In this work, we focus on the $\sim$5\,Myr old \citep{monnier2019} quadruple system HD\,34700 \citep[HIP\,24855; d=350.9$\pm2.4$\,pc;][]{Gaia_dr3_2022}. It consists of a close ($a_{\rm bin}$=0.69\,au) near-equal-mass Herbig Ae spectroscopic binary ($M_{\rm bin}=4M_\odot$, HD\,34700 AaAb; hereafter, \sys) and two low-mass M-type companions located at distances of $5\farcs2$ (B, $\sim0.6M_\odot$) and $9\farcs3$ (C, $\sim0.4M_\odot$; \citealt[][]{sterzik2005}). The system was first observed by \cite{monnier2019} in near-infrared (NIR) scattered light with Gemini/GPI. They found that the binary is surrounded by a disk with a cavity and multiple substructures. As shown in Fig.\,\ref{fig:overview}a, it displays a large inner dust-depleted cavity surrounded by a wide elliptic ring-like feature at $R=$175\,au, from which at least six spirals with extensive opening angles ($\sim$30$^\circ$ to 50$^\circ$) originate. Interestingly, these observations revealed a discontinuity toward the north of the ring, which was later confirmed by Subaru/SCExAO and SPHERE observations \citep{uyama2020, Columba_ea_2024}, together with several shadows on the ring and spirals, suggesting the presence of an inclined inner disk or a warp. Recent observations further detected a symmetric ring in H$\alpha$ emission ($R=65-120$\,au) inside the asymmetric IR ring (Fig.\,\ref{fig:overview}\,b), tracing sub-$\mu$m-sized dust particles \citep{Columba_ea_2024}. Low-resolution ($\lesssim0\farcs35$) Submillimetre Array (SMA) and ALMA observations revealed a prominent azimuthal asymmetry in the mm continuum, located at the edge of the scattered-light cavity \citep{Benac_ea_2020, Columba_ea_2024}. 

Continuum asymmetries can arise from azimuthally trapped dust particles \citep{Birnstiel2013, Bae2016}. This trapping occurs in local pressure maxima, which could develop due to anticyclonic vortices caused by the Rossby wave instability \citep[RWI, ][]{Lovelace1999, Li_ea_2000}. For the RWI to be triggered, the disk must show strong gas density gradients, which tend to be present at the edges of gas cavities \citep[e.g.,][]{Bracco_ea_1999}. Other favorable mechanisms include dead zones \citep{Flock_ea_2015, Flock_ea_2017a} or the anisotropic infall of material \citep{Bae2015, Kuznetsova2022}. To sustain the RWI and observe long-lived anticyclonic vortices, the turbulence within the disk must remain low \citep[$\alpha\lesssim10^{-3}$, ][]{Li_ea_2000}. Besides local over-densities in dust rings, pronounced dust continuum crescents have only been observed in a few transition disks, such as IRS\,48 \citep{vanderMarel_ea_2013} and HD\,142527 \citep{Casassus2013}. Although several attempts have been made to characterize gas motions around these crescents and to determine whether they can be explained by vortices \citep{robert_ea_2020, Boehler_ea_2021, vanderMarel_ea_2021a, Woelfer_ea_2025}, the results have thus far remained inconclusive due to the complexity of the disk kinematics and the limited resolution of these observations.

The multiple IR spirals in HD\,34700A and their varying opening angles remain challenging to explain. Gravitational instability and stellar flybys have been considered unlikely \citep[][respectively]{monnier2019, shuai2022}. Additionally, the binary separation is too small ($<1$\,au) to allow for any periodic perturbation over its orbit, which would be needed to drive the spirals. A possible explanation for the origin of the observed features is the presence of a massive companion of a few Jupiter masses located within the cavity, along with the system's interaction with its surrounding environment \citep{monnier2019, Columba_ea_2024}. Recent disk observations have reported some of the effects of late-stage infall of ambient material from the molecular cloud \citep[e.g.,][]{Garufi_ea_2024, gupta2023, Huang_ea_2020, Huang_ea_2021, Huang_ea_2023, Speedie_ea_2025} in disks that show extended spiral features. Theoretical studies show that these processes can initiate and sustain spiral arms for several thousand years after the infall event \citep{Dong_ea_2016, Calcino_ea_2025, kuffmeier2020} and can also lead to disk warping \citep[e.g.,][]{Akeson_ea_2007, Dullemond_ea_2019, Kuffmeier_ea_2021, Young_ea_2022}.

This paper presents new ALMA Band 6 continuum and CO molecular line observations at high-resolution toward HD\,34700 and is structured as follows. In Sect. \ref{sec:calib} we describe the observations, applied calibration, and imaging choices. Section \ref{sec:method} explains the methodology used to fit the CO kinematic data. Our results are presented in Sect. \ref{sec:results}, with subsections on the continuum crescent, molecular line emission and kinematics, and line spectra analysis. Finally, we discuss our findings in Sect. \ref{sec:discussion} and state our conclusions in Sect. \ref{sec:conclusion}.

%%%%%%%%%%%%%%%%%%%%%%%%%%%%%%%%%%%%%%%%
%%%%%%%%%%%%%%%%%%%%%%%%%%%%%%%%%%%%%%%%

%%%%%%%%%%%%%%%%%%%%%%%%%%%%%%%%%%%%%%%%%%%%%%
\section{Observations, calibration, and imaging}
%%%%%%%%%%%%%%%%%%%%%%%%%%%%%%%%%%%%%%%%%%%%%%
\label{sec:calib}
We present new ALMA Band~6 observations of the quadruple system HD\,34700 (2022.1.00760.S; PI:~Stadler) with two short-baseline (SB) and eleven long-baseline (LB) execution blocks (EBs), with total on-source time of 1.6\,h and 8.6\,h, respectively. The SB observations in configuration C-3 were taken on October 9, 2022 with a mean precipitable water vapor (PWV) of 0.3 mm. The LB executions in C-6 spanned three weeks with a mean PWV in the range of 0.3 to 1.4\,mm. Details on the EB measurement sets are listed in Table~\ref{tab:observations}. The spectral setup was designed to study the kinematics of the circumbinary disk. Thus, one spectral window (SPW) was centered on the \twCOfull line at a native spectral resolution of 92\msec{} (70.6\,kHz) and one around the \thCOfull and \eiCOfull molecular lines, both at 192\msec{} (141\,kHz). The remaining two SPWs target the continuum emission centered at 218.0 and 232.6 GHz, each with a bandwidth of 1.9\,GHz and native frequency spacing of 1.13\,MHz ($\sim$1.5\kmsec{} native channel spacing).

The data calibration and imaging were performed using \texttt{CASA v.6.2.1} \citep{CASA_Team_ea_2022} following the procedure of the exoALMA collaboration presented in \cite{loomis2025}. This pipeline was designed explicitly for high angular and spectral resolution kinematical data. Starting with the pipeline-calibrated data, we applied self-calibration (self-cal) to all of our EBs. First, we flagged any prominent line emission detected in each SPW and then spectrally averaged the EBs by a factor of 12. This pre-averaging did not show any significant negative effect on the final obtained images as shown in \cite{curone2025}. Following the exoALMA pipeline, we performed one phase-only self-calibration round with solution intervals of \texttt{solint}=`inf'  on each EB, before spatially aligning and concatenating the individual SB and LB EBs. We then conducted three rounds of phase-only self-cal (\texttt{solint}=[`inf', 340s, 120s]) on the concatenated SB observations and combined them with the LB EBs. Three additional rounds of phase self-cal (\texttt{solint}=[`inf', 340s, 120s]) and one round of amplitude and phase self-cal (\texttt{solint}=`inf') were performed on the concatenated dataset, increasing the final continuum signal-to-noise ratio (S/N) by over 30\% compared to the concatenated non-calibrated dataset. These self-calibration tables were then applied to the whole data set, including the molecular lines. Finally, a continuum subtraction was performed before splitting out the final line emission measurement set (ms) tables.

The continuum was imaged using two manually adapted ellipses: one centered on \sys{} with a semi-major axis of $1\farcs2$ and an inclination of 40\degree{} and a second centered on HD\,34700B with a radius of $0\farcs5$ and an inclination of 26\degree{}. For the molecular line imaging, we followed the procedure outlined in \citet[see their Fig. 10]{loomis2025}.

The continuum images presented in this paper were imaged CLEANing down to 3 sigma and with robust parameters (rob.) of 0.5 and $-1.5$ \citep{Briggs_1995}, corresponding to synthesized beams of $137\times115$ mas$^{2}$ (86.3°) and $92\times66$ mas$^{2}$ (87.3°), respectively. The resulting RMS noise levels are 5.1\,µJy\,beam$^\mathrm{-1}$ (rob. 0.5) and 22.7\,µJy\,beam$^\mathrm{-1}$ (rob.$\,-1.5$). We measured the peak continuum intensity for the robust 0.5 image to be 2.32\,mJy\,beam$^\mathrm{-1}$ (S/N of 458\,$\sigma$) with a total flux density of 8.42$\pm0.8$\,mJy (10\% absolute flux error in ALMA Band 6) within the CLEAN mask, in line with previous measurements \citep{Benac_ea_2020, Columba_ea_2024}.

The CO molecular lines were CLEANed down to a threshold of $3\,\sigma$ and then imaged using a robust parameter of 1.5 to obtain the ``fiducial'' high-sensitivity images and a robust parameter of 0.5 to obtain ``high-resolution'' images. These robust values result in synthesized beams of $0\farcs22\times0\farcs17$ (96.0\degree{}) and $0\farcs14\times0\farcs11$ (88.8\degree{}). For the \twCOfull image cubes, we measured an RMS noise outside the CLEAN mask of 0.99\,mJy\,beam$^\mathrm{-1}$ (rob.\,1.5, RMS noise in the Rayleigh-Jeans approximation: 0.62\,K) and 1.12\,mJy\,beam$^\mathrm{-1}$ (rob.\,0.5, 1.65\,K) both with 100\,\msec{} channels. The \thCOfull and \eiCOfull emission lines were imaged with 200\,\msec{} channels and have noise levels of [0.74, 0.85]\,mJy\,beam$^\mathrm{-1}$ ([0.46, 1.2]\,K) and [0.59, 0.68]\,mJy\,beam$^\mathrm{-1}$ ([0.38, 1.0]\,K) on the robust [1.5, 0.5] cubes, respectively. An overview of the above-mentioned imaging parameters, RMS noises, and flux densities, for the continuum and CO molecular lines can be found in Table\,\ref{tab:imaging_params}.

In the following, we  solely focus on analyzing the circumbinary disk \sys{}. The observations and analysis of the circumstellar disk around binary companion HD\,34700B are presented in Appendix\,\ref{app:hd34700b}. We  note that we did not detect any significant ($>3\sigma$) continuum or molecular line emission at the positions of either HD\,34700C or proposed companion D \citep[$6\farcs45$ along PA$\approx-61^\circ$][]{monnier2019}, nor any emission that bridges over \sys{} and B. 

%%%%%%%%%%%%%%%%%%%%%%%%%%%%%%%%%%%%%%%%%%%%%%
\section{Methodology}
\label{sec:method}
%%%%%%%%%%%%%%%%%%%%%%%%%%%%%%%%%%%%%%%%%%%%%%
We employed the \verb|discminer| code package \citep{Izquierdo_ea_2021} to model the molecular line emission channel-by-channel and analyze the disk gas kinematics. This state-of-the-art code fits a Keplerian model to the local line profile of each spatial pixel $i$ and velocity channel, $j,$ that maximizes the log-likelihood function
\begin{equation} \label{eq:loglikelihood}
 \log \mathcal{L} = -0.5\sum_{j}^{n_{\rm ch}} \sum_{i}^{n_{\rm pix}}  \left[I_{\rm m}(r_i, \upsilon_j) - I_{\rm d}(r_i, \upsilon_j)\right]^2/\sigma_{i}^{2} ,    
\end{equation}

\noindent between the observed $I_{\rm d}$ and model intensity $I_{\rm m}$. Each pixel is weighted by a factor, $\sigma_i$, with the standard deviation
 of its intensity measured in line-free channels, analogously to the RMS noise. To obtain the best-fit model parameters, the code utilizes Markov chain Monte Carlo (MCMC) ensemble sampler \textsc{emcee} \citep{Emcee_2019} to sample the parameter posterior distribution. The model is parametrized in the radial direction using a double-power law for the peak intensity and a monotonically decreasing one for the linewidth. The disk surface height is described by an exponentially tapered power law, which is assumed to be infinitely narrow, and the velocity is a Keplerian profile corrected for the surface heights $\upsilon_\mathrm{K}(R,z)$ \citep[see][for functional forms]{izquierdo2025}.

We fit each molecular line tracer independently, employing the high-resolution cubes for \twCO{} and \thCO{} and fiducial cube for \eiCO{}. For the initial parameter guesses in each run, we adopted the literature values for combined stellar mass of the spectroscopic binary ($M_\star=4.0\,M_\odot$) and inclination ($i_0=40\degree{}$; \citealt{torres2004, Columba_ea_2024}). For the remaining parameters, including position angle PA, systemic velocity $v_\mathrm{LSRK}$, and power-law profiles for the intensity, line-width, and line-slope, we estimated the starting values using a preliminary \verb|discminer| model prototype applied to the data channels. We did not prescribe a back surface in the model, as the channel maps and line profiles did not show any contribution from it \citep[see][for a detailed discussion]{izquierdo2025}. In other words, we only observe the disk's near-side. For a disk seen at a moderate inclination of $i=40\degree{}$, this hints toward a relatively flat disk emission height. Furthermore, we masked the innermost beam size in diameter from the disk center due to cavity and beam smearing effects. We employed 200 walkers for 21 parameters and ran the fits for 37,000 (\twCO{}), 20,000 (\thCO{}), and 15,000 (\eiCO{}) steps until the marginalized posteriors were reasonably well converged and Gaussian. The best-fit model parameters are the 16th and 84th percentiles of the posterior distributions in the final 10\% of the walkers, as shown in Table\,\ref{tab:params}.

%%%%%%%%%%%%%%%%%%%%%%%%%%%%%%%%%%%%%%%%%%%%%%
\section{Results}
\label{sec:results}
%%%%%%%%%%%%%%%%%%%%%%%%%%%%%%%%%%%%%%%%%%%%%%
\subsection{Dust images}
%%%%%%%%%%%%%%%%%%%%%%%%%%%%%%%%%%%%%%%%%%%%%%
Figure\,\ref{fig:overview} presents a multi-wavelength observational overview of \sys{}. Panel (a) shows the infrared (IR) VLT/SPHERE polarized scattered-light observations and (b) the VLT/ZIMPOL H$\alpha$ observations of the circumbinary disk, both from \cite{Columba_ea_2024}. The IR observations exhibit an inner cavity surrounded by a prominent asymmetric ring at $R\approx0\farcs50$. Several spiral arms originate from this ring, and it displays an apparent discontinuity or ``break'' in the north. Inside the IR ring, the H$\alpha$-observations display another ring extending from 186 to 335\,mas, which traces sub-micron-sized particles \citep{Columba_ea_2024}. The $m=2$ spirals beyond the H$\alpha$-ring align with the IR ring, suggesting the IR ring is actually the spirals, albeit at lower angular resolution.

Our new ALMA 225.3\,GHz continuum observations in panel (c) reveal a highly asymmetric dust crescent co-located with the IR ring at $0\farcs39$ (138\,au) and an inner unresolved dust disk emission of 55\,µJy\,beam$^\mathrm{-1}$ ($\sim11\sigma$), measured by enclosing one beam-size centered on the binary. The continuum emission morphology resembles dust trapped in an anticyclonic vortex \citep{Birnstiel2013}, which is smeared out over the disk's counterclockwise rotation. A thorough analysis of the continuum emission will be presented in a forthcoming paper \citep{Fasano_ea_2026}. A gallery exhibiting the dust emission morphologies is presented in Fig.\,\ref{fig:highres_dust_images}.

%%%%%%%%%%%%%%%%%%%%%%%%%%%%%%%%%%%%%%%%%%%%%%
\subsection{CO intensity features}
\label{sec:line_emission}
%%%%%%%%%%%%%%%%%%%%%%%%%%%%%%%%%%%%%%%%%%%%%%
The \twCOfull{}, \thCOfull{}, and \eiCOfull{} zeroth and peak intensity (eighth) moment maps using the robust 1.5 cubes shown in the middle and lower row of Fig.\,\ref{fig:overview}, respectively. These moment maps were computed using \verb|bettermoments| \citep{bettermoments}, with a $2\sigma$ clip applied to the data. The peak line emission of \twCO{}, \thCO{}, and \eiCO{} are [60.9, 32.7, 10.9]\,mJy\,beam$^{-1}$, respectively. All three CO isotopologs exhibit an inner cavity in their emission, with its elliptic shape elongated toward the north-south direction and, thus, it is asymmetric with respect to the disk major axis. We expected \eiCO{} to be more optically thin, which would serve as an indication that it is tracing not only the temperature, but also the density variations as well; this, in turn,  suggests the presence of a real gas cavity. However, the observed size of the gas cavity ($R\sim40\,$au) cannot be explained by the (tidal) truncation of the inner eccentric binary. Its maximal separation is a mere 0.7\,au \citep[$e_\mathrm{bin}=0.25$,][]{torres2004}; hence, it would  only have the capacity to carve out a cavity of a maximum size  $\leq6a_\mathrm{bin}=4.2\,$au \citep{Penzlin_ea_2024}. 

%%%%%%%%%%%%%%%%%%%%%%%%%%%%%%%%%%%%%%%%
\begin{figure}[t]
    \centering
    \includegraphics[width=0.8\linewidth]{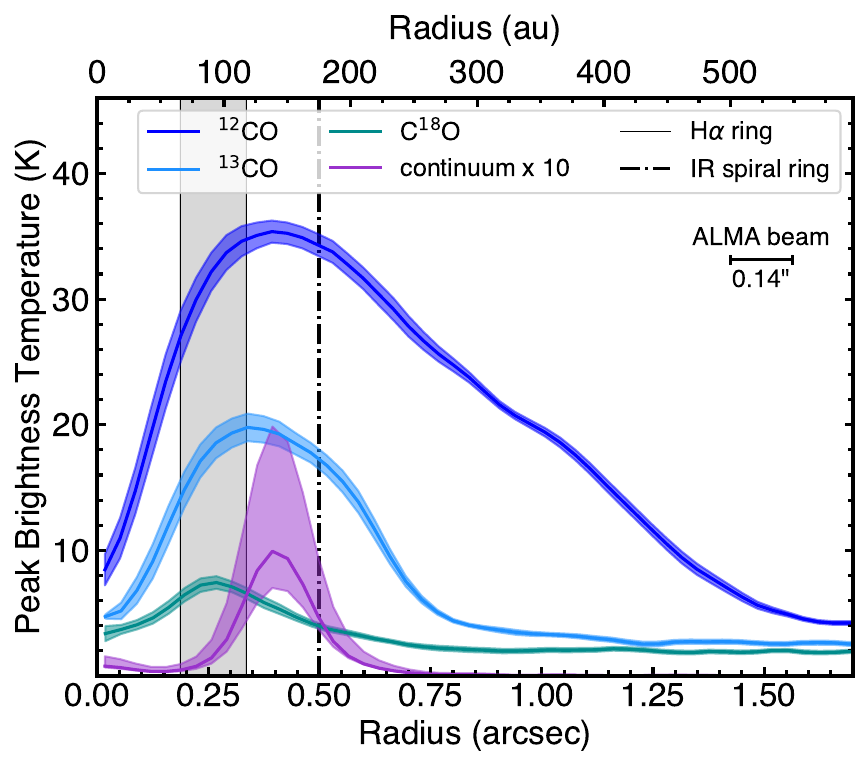}
    \caption{Brightness temperature radial profiles of CO isotopologs and dust continuum using the high-resolution cubes $0\farcs14\times0\farcs11$ (49\,x\,39\,au) cubes. The latter was multiplied by a factor of 10 to enhance visibility. The colored shaded region shows the standard deviation within each annulus. For the asymmetric continuum emission this traces the pronounced azimuthal variations, rather than an error. The gray-shaded area shows the extent of the H$\alpha$ ring, and the dashed-dotted line shows the peak of the IR ring \citep[values from][respectively]{Columba_ea_2024, uyama2020}. }
    \label{fig:rad_prof}
\end{figure} 
%%%%%%%%%%%%%%%%%%%%%%%%%%%%%%%%%%%%%%%%

We used the \verb|GoFish| package \citep{Teague_2019_gofish} to deproject and azimuthally average the high-resolution peak intensity and continuum maps employing the \thCO{} geometrical parameters from \verb|discminer|. We made this choice to ensure a coherent analysis, as the individual best-fit geometrical parameters for the CO isotopologs differ significantly (see Table\,\ref{tab:params}). We regard the inferred geometry for \thCO{} as the most reliable because it probes deeper into the disk vertically compared to \twCO{}, making it significantly less dynamically perturbed. Additionally, it exhibits a higher S/N than \eiCO{}, which allows for a more robust fit to the data. For the continuum profile, we further mask the azimuthal regions devoid of emission, which are [$-90,+90$]\degree in the disk frame measured from the redshifted major axis. We show the radial emission profiles using the Rayleigh-Jeans approximation to convert to units of K in Fig.\,\ref{fig:rad_prof}. The gas cavity is seen across all tracers and its edge is at the inner radius of the H$\alpha$-ring at $R=0\farcs2$ (65\,au) inferred in \cite{Columba_ea_2024}, as it aligns with the full-width at half maximum (FWHM) of all CO line emission peaks. Interestingly, the \eiCO{} emission peak is co-located at the center of the H$\alpha$ ring. 

\begin{figure}[t]
    \centering
    \includegraphics[width=1\linewidth]{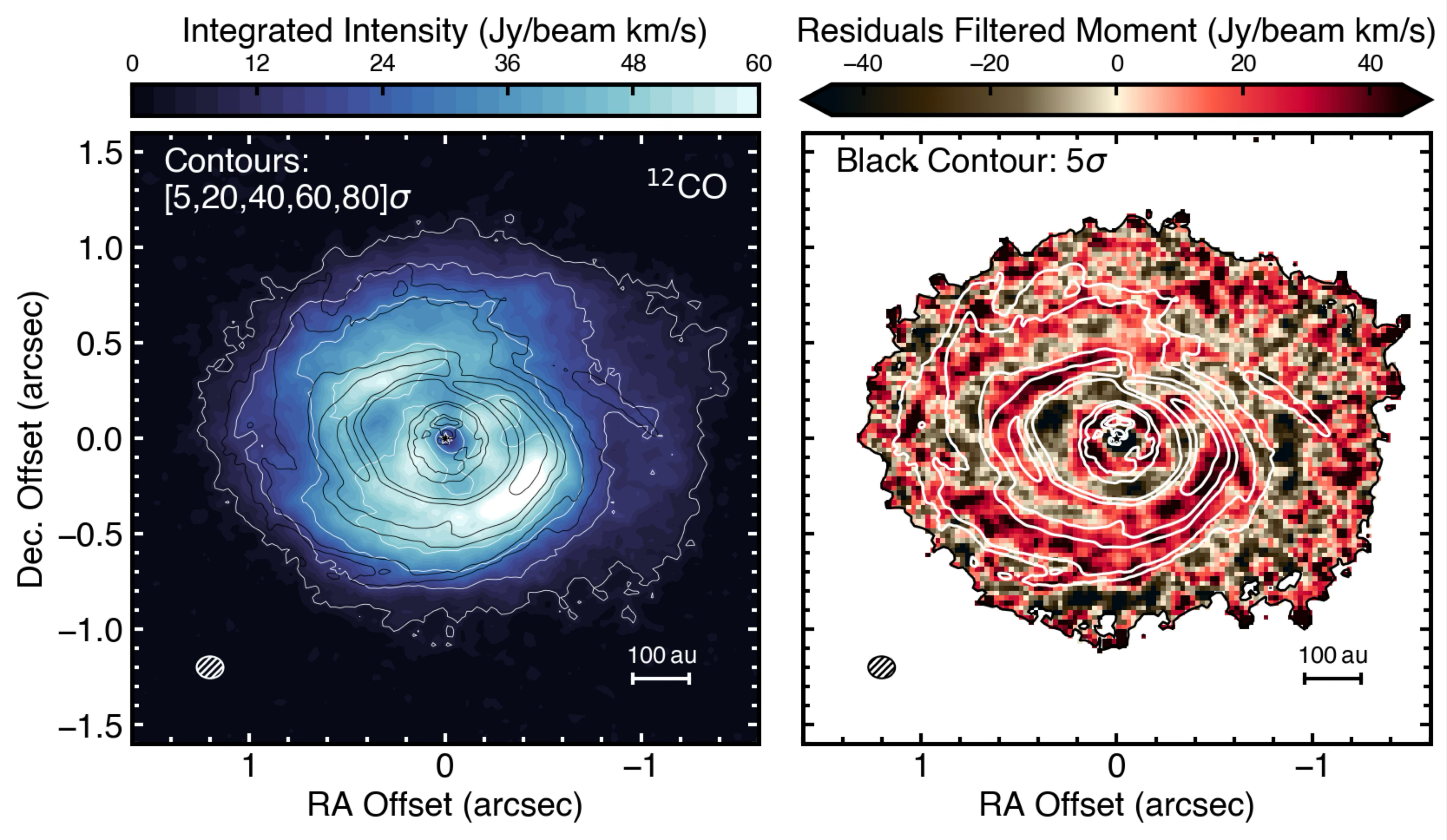}
    \caption{Line emission features of \twCO{}. Left: Moment 0 map using the high-resolution cube. Right: Residuals from a moment 0 map subtracted with a high-pass filtered moment map, limited to disk emission above 5\,$\sigma$. The black and white contours in both maps highlight the IR emission, identical to Fig.\,\ref{fig:overview}\,(a).}
    \label{fig:mom0_12co_fitler}
\end{figure}

The \twCO{} peak intensity map in Fig.\,\ref{fig:overview}~(g) displays an arc-shaped feature spanning several beam sizes toward the northeast, broadly aligned with a faint spiral arm detected in H-band $Q_\phi$ \citep[C1;][]{Columba_ea_2024}. Overall, the \twCO{} emission closely follows the contours of the IR ring, as seen in the \twCO{} channel maps in Fig.\,\ref{fig:12co_channels} and the moment 0 map shown in the left panel of Fig.\,\ref{fig:mom0_12co_fitler}. We apply a high-pass filter to the moment 0 map, with the following kernel $\omega(R) = \omega_0 (R/1^{\prime\prime})^\gamma$ expanding with radius, $R$, defined in \cite{Speedie_ea_2024}. We adopted a power-law index of \(\gamma = 0.1\) and a kernel width of \(\omega_0 = 3\) pixels. The filtered map is then subtracted from the original moment 0 map and the resulting residuals are shown in the right column of Fig.\,\ref{fig:mom0_12co_fitler}. The spatial correlation between positive \twCO{} filtered residuals and the IR ring is striking. It is also noteworthy that there is an asymmetry in the CO emission between the northern and southern sides of the disk; the southern side is brighter, despite it being the side that points away from us \citep[e.g.,][]{monnier2019}. The southern IR spiral-arc structure corresponds to positive residuals across nearly half of the disk’s azimuth. This connection  is discussed in more detail in the following sections.

%%%%%%%%%%%%%%%%%%%%%%%%%%%%%%%%%%%%%%%%
%%%%%%%%%%%%%%%%%%%%%%%%%%%%%%%%%%%%%%%%

%%%%%%%%%%%%%%%%%%%%%%%%%%%%%%%%%%%%%%%%%%%%%%
\subsection{Gas kinematical diagnostics}
\label{sec:kinematics}
%%%%%%%%%%%%%%%%%%%%%%%%%%%%%%%%%%%%%%%%%%%%%%
\subsubsection{Channel maps} \label{sec:ch_maps}
\begin{figure}[h]
    \centering
    \includegraphics[width=1\linewidth]{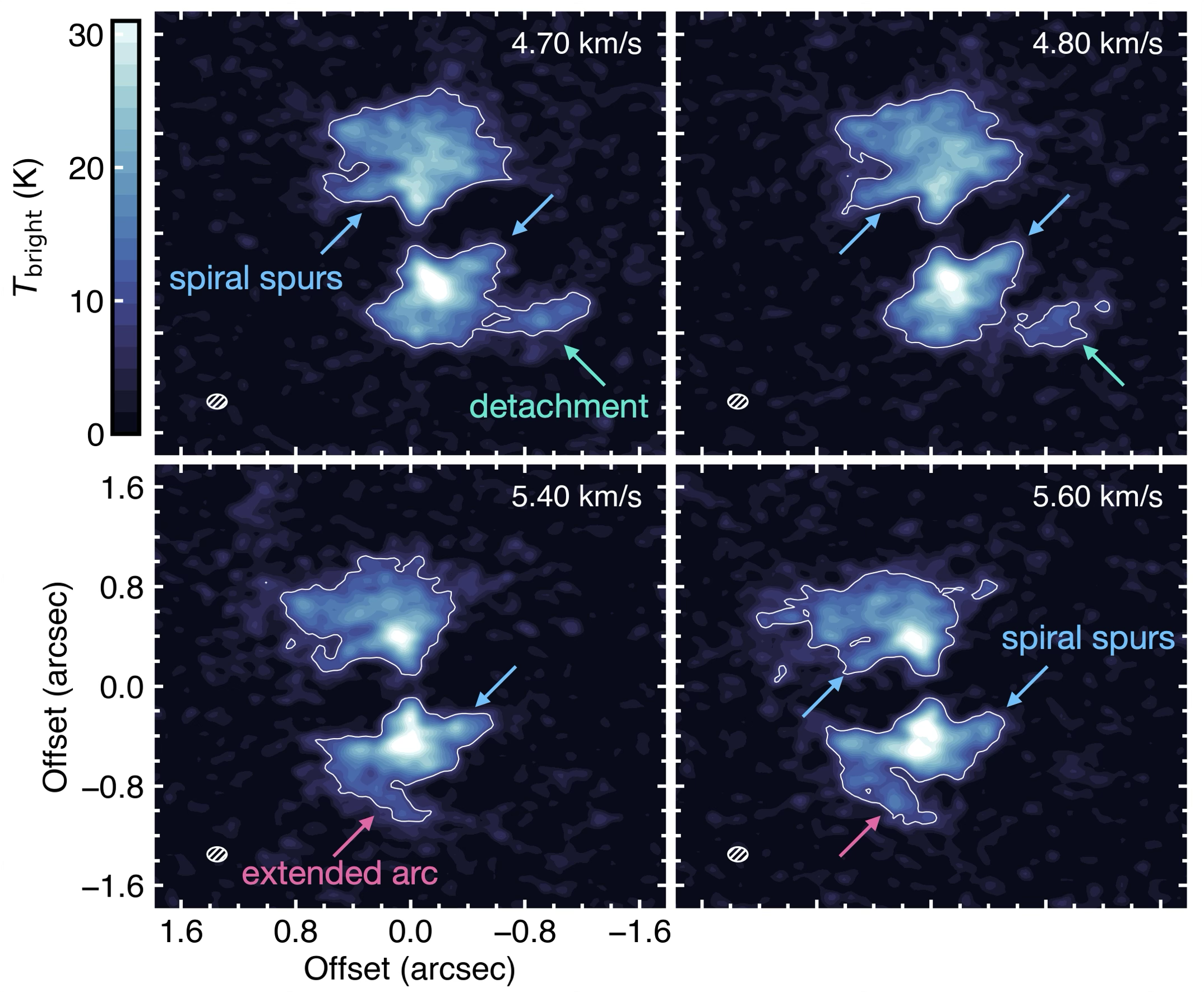}
    \caption{Selected \twCO{} channel maps with highlighted non-Keplerian features. White contours enclose 5$\sigma$. The beam size of the high-resolution cube employed is shown in the lower left corner and channel id in the upper right corner.}
    \label{fig:12co_select_chan}
\end{figure}

%%%%%%%%%%%%%%%%%%%%%%%%%%%%%%%%%%%%%%%%%%%%%%
In the appendix, we present selected channel maps for the CO isotopologs in Figs. \ref{fig:12co_channels}, \ref{fig:13co_channels}, and \ref{fig:c18o_channels}. Again, we overplotted the scattered light contours in the \twCO{} channel-map gallery. The channel emission exhibits substantial deviations from circular Keplerian rotation, manifesting as spurs that align with the IR ring contours. Along the spiral arms, near the minor axis of the disk, the line emission shows strong vertical (or radial) flows, which we are particularly sensitive to in our  (LoS). This causes the emission to shift and spread into adjacent channels, most pronounced on the southern side of the disk. In addition, some of the emission even detaches from the main southern channel wing in channels $4.7-5.1$\,\kmsec{} seen in Fig.\,\ref{fig:12co_select_chan}. In the following channels ($5.3-5.7$\,\kmsec{}), an extended arc emerges, which is attached to the eastern and southern emission, as well as another spiral feature that now blueshifts the southern emission to the east ($5.7-5.9$\,\kmsec{}). In contrast, the \thCO{} and \eiCO{} channel maps exhibit a smoother emission morphology.

%%%%%%%%%%%%%%%%%%%%%%%%%%%%%%%%%%%%%%%%
%%%%%%%%%%%%%%%%%%%%%%%%%%%%%%%%%%%%%%%%

This morphological contrast between \twCO{} and \thCO{} and \eiCO{} emission can also be seen in the CO channel maps gallery of Figs.\,\ref{fig:discminer_channels_12co}$-$\ref{fig:discminer_channels_c18o}, showing data, \verb|discminer| model, and residuals channels together with their peak intensity maps. The complexity of the \twCO{} emission cannot be captured by the pure Keplerian model, while for \thCO{} and \eiCO{}, the \verb|discminer| model describes the emission morphology well. 

\begin{figure}[t]
    \centering
    \includegraphics[width=\linewidth]{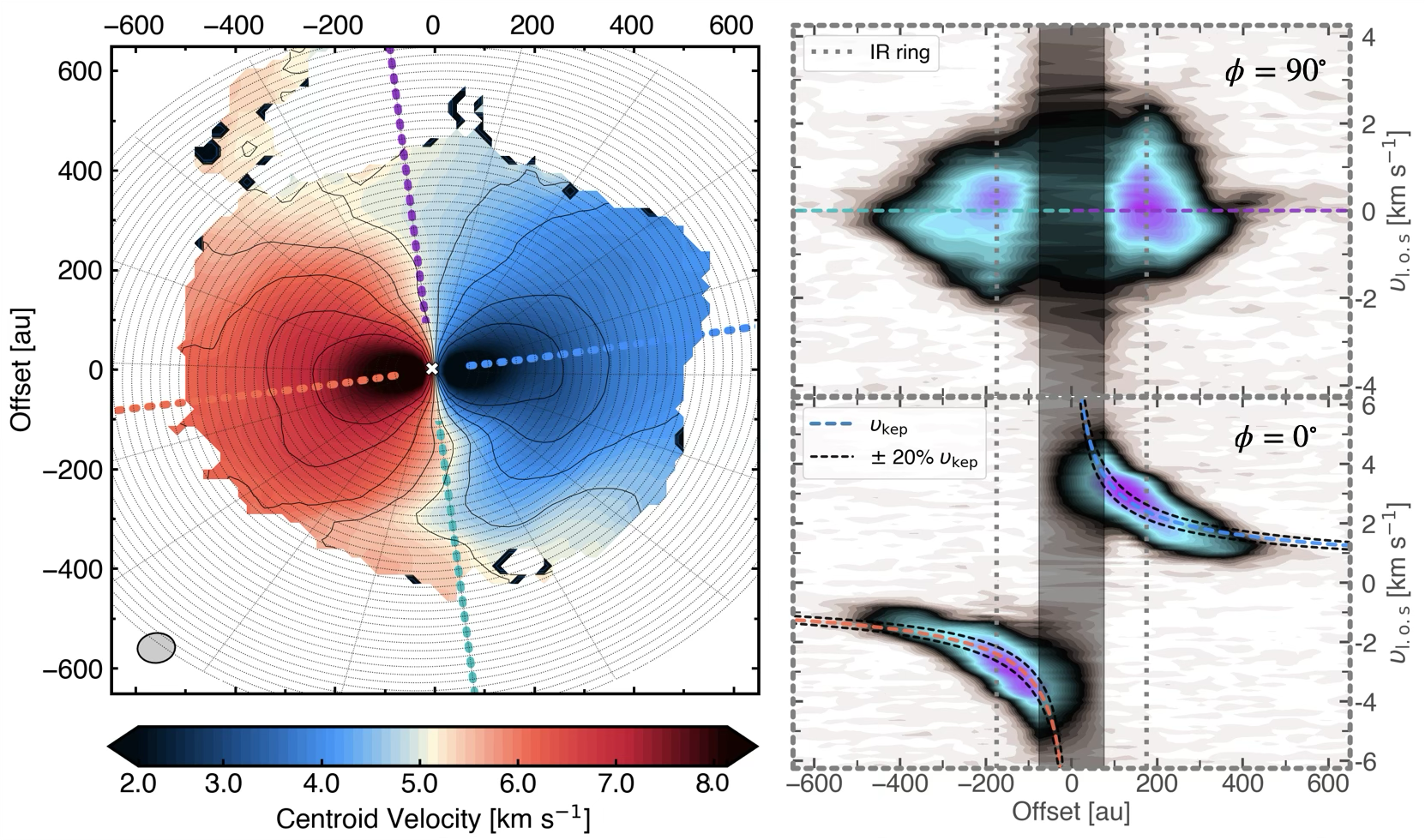}
    \caption{\twCO{} position-velocity diagram for fiducial cube. Spectra are extracted along all four angles of the major (red and blue at $\phi=$0\degree{}) and minor (cyan and purple at $\phi=$90\degree{}) disk axes, as highlighted in the Gaussian centroid moment map (left panel). The dashed lines in the PV diagrams (colors corresponding to axes on the left) highlight the disk's Keplerian rotation. The gray shaded areas mask one major beam size in radius from the center. In the $\phi=0$ panel, the enclosing black dashed lines correspond to 20\% deviations from Keplerian rotation. The location of the IR ring ($R=175\,$au) is plotted in gray dotted lines in both PV diagrams.}
    \label{fig:PV_diagram}
\end{figure}

To further understand the prominent \twCO{} kinematic features, we utilized the inferred \twCO{} \verb|discminer| geometry (Table \,\ref{tab:params}) to deproject and plot the position-velocity diagrams for \twCO{} along the disk axes, as shown in Fig.\,\ref{fig:PV_diagram}. In the panel displaying the emission along the major axis (\(\phi=0^\circ\)), we also overlayed the expected Keplerian rotation and 20\% deviations for comparison. The emission is significantly spread out beyond those limits in velocity space (at approximately $\sim 3\,$\kmsec{}). However, typical deviations from Keplerian rotation are expected to be on the order of a few percent \citep[i.e., $0.1\times\upsilon_\mathrm{kep}(100\,\rm{au})\approx0.4\,$\kmsec{}; e.g.,][]{stadler2025}. The broadening of the lines is especially noticeable along the IR ring and its interior. The Keplerian shear within one major beam size (75\,au) at this location ($R=175$\,au) can reach a maximum of $\Delta\upsilon_{\rm kep}\approx1$\,\kmsec{}, which cannot fully account for the observed broadening. The spectral cut along the disk's minor axis ($\phi=90$\degree{}) reveals spread-out emission at radii interior to $\lesssim300\,$au. These broad-line widths likely result from rapid and varied radial or vertical motions, which significantly contribute to the LoS velocity (especially at the minor axis). Simultaneously, beam-smearing effects within a few beam sizes from the center can lead to artificial sub-Keplerian rotation and the spreading of emission \citep[e.g.,][]{Boehler_ea_2021}.

%%%%%%%%%%%%%%%%%%%%%%%%%%%%%%%%%%%%%%%%
%%%%%%%%%%%%%%%%%%%%%%%%%%%%%%%%%%%%%%%%

%%%%%%%%%%%%%%%%%%%%%%%%%%%%%%%%%%%%%%%%%%%%%%
\subsubsection{CO spectral analysis}
\label{sec:spectra_12co}
%%%%%%%%%%%%%%%%%%%%%%%%%%%%%%%%%%%%%%%%%%%%%%

%%%%%%%%%%%%%%%%%%%%%%%%%%%%%%%%%%%%%%%%
\begin{figure*}[t!]
    \centering
    \includegraphics[width=0.85\linewidth]{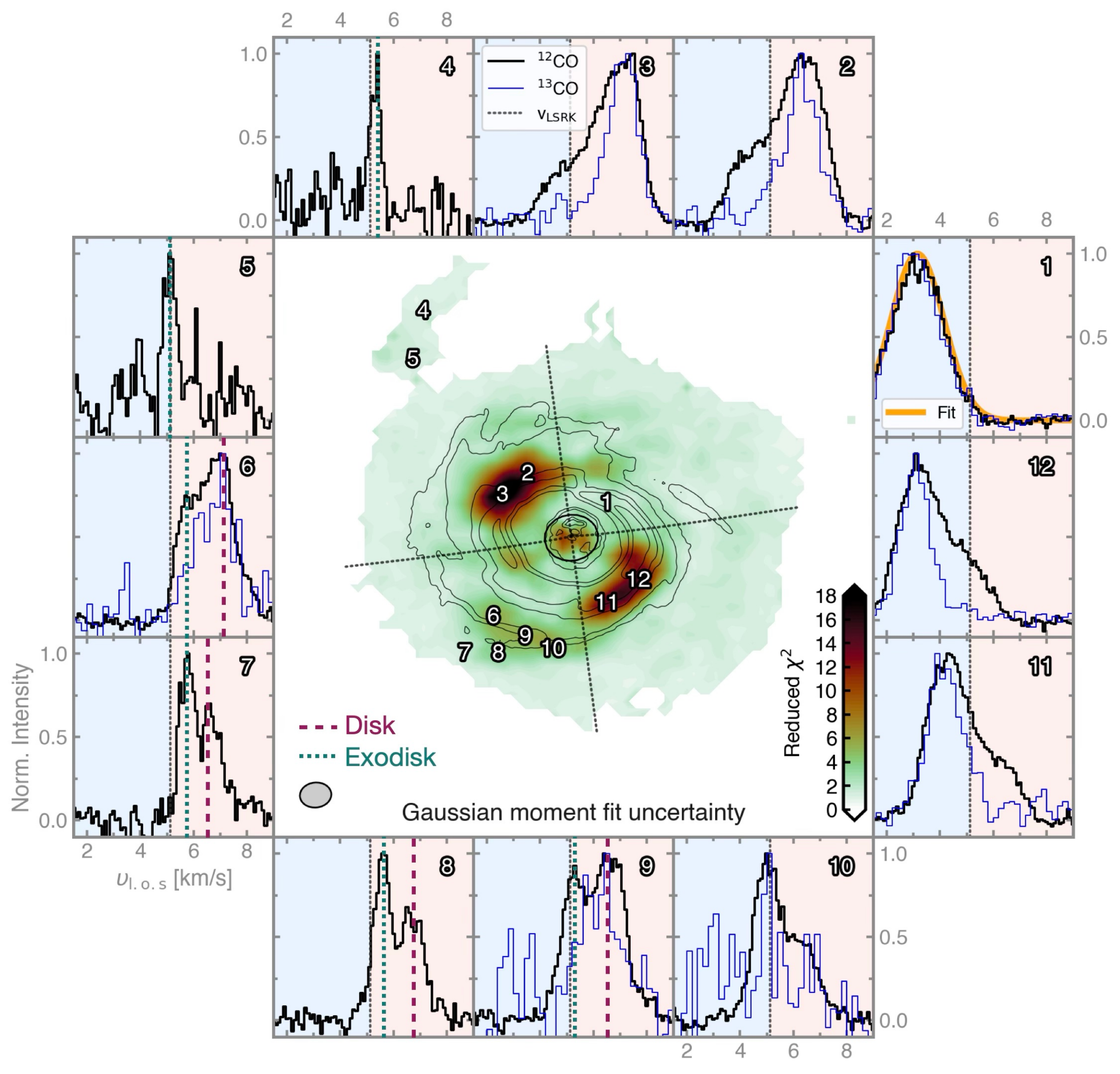}
    \caption{Line spectra diagnostics. Selected \twCO{} (black) and \thCO{} (blue) spectra normalized to their peak extracted at disk locations marked in the center map. The map shows the reduced-$\chi^2$ values (Eq.\,\ref{eq:chi2}) from the Gaussian moment map fit to the line profile (see accurate fit in panel 1), masked at $5\sigma$. High uncertainties indicate non-Gaussian-shaped line profiles. The line profiles are split by the $v_\mathrm{LSRK}$ into regions that are either blue- or redshifted to the systemic velocity. Pixels 6-9 that exhibit secondary components in their spectra that are tracing an ``exodisk'' line component (cyan vertical lines), in addition to the underlying rotating disk component (purple lines). The black contours show the IR scattered light as in Fig.\,\ref{fig:overview}~(a) and the dotted lines highlight the disk's minor and major axes. }
    \label{fig:line_profiles}
\end{figure*} 
%%%%%%%%%%%%%%%%%%%%%%%%%%%%%%%%%%%%%%%%

In this subsection, we describe our line profile analysis of the \twCO{} and \thCO{} spectra, aimed at  gaining further insights into the different velocity structures. First, within the \verb|discminer| analysis framework, we fit a single Gaussian function to the line profile of each pixel to create moment maps for both the data and model cubes, which then trace the line's peak intensity, centroid LoS velocity, and the Gaussian width ($n_\mathrm{pars}=3$). In the center panel of Fig.\,\ref{fig:line_profiles}, we present the reduced-$\chi^2$ uncertainty associated with the fit $I_{\rm{mom}, j}$ to the fiducial \twCO{} data cube calculated via

\begin{equation} \label{eq:chi2}
\chi_\nu^2=\left(n_{\mathrm{ch}}-n_{\mathrm{pars}}\right)^{-1} \sum_j^{n_{\mathrm{ch}}}\left(I_{\rm{d}, j}-I_{\rm{mom}, j}\right)^2 / s_j^2,
\end{equation}

\noindent where $s_j$ is a uniform weighting factor that is the standard deviation of the observed intensity per channel $I_{\rm{d}, j}$, measured from line-free pixels \citep{izquierdo2025}.

Dark red colors highlight regions with high fitting uncertainty, where a single Gaussian function provides a poor description of the local line profile. In other words, these regions show the blend of multiple (spatially) unresolved velocity flows. We extract the local \twCO{} line profiles at these locations, highlighted by the numbers overplotted on the $\chi^2$-map. The \thCO{} spectra are shown to facilitate comparison with the underlying disk rotation (panels 1-3, 6, and 9-12). Panel 1 illustrates a spectrum with low fitting uncertainty, indicating that in these regions, a Gaussian fit (orange line) provides an accurate description of the data.

The map shows a local increase in the reduced-$\chi^2$ in the southeastern part of the disk (panels 6-10). It reveals a secondary component in the line spectra (left peak), offset by over 1\,\kmsec{} in its LoS velocity ($\upsilon_\mathrm{los}$) from the disk component (right peak). This additional component cannot be due to the disk's backside emission, since it should not be visible on the southern side, pointing away from us. We note the spatial co-location of these features with the southeastern IR spirals. The map further displays prominent fitting uncertainties in the disk regions (panels 2, 3, 11, and 12 at $R\approx190\pm10\,$au) co-located with the asymmetric IR ring and H$\alpha$ spirals. At these locations, the line spectra exhibit shoulders in addition to the underlying Keplerian rotation outlined by the \thCO{} spectra. Lastly, the spectra in panels 4 and 5 show the motion of the arc-shaped feature in the northeast. While this feature does not appear to move at smaller radii, the spectrum at pixel 4 exhibits a slight redshift (by a few hundred \msec{}) in the \(v_\mathrm{LSRK}\), indicating that this structure is moving away from us, in line with the expected disk rotation in these regions. In the following discussion, we discuss our  interpretation of these secondary line components and shoulders as ``exodisk'' \citep[following][]{Speedie_ea_2025} material detached from the disk, which is infalling toward the disk in our LoS (redshifted w.r.t. $\upsilon_\mathrm{LSRK}$).

%%%%%%%%%%%%%%%%%%%%%%%%%%%%%%%%%%%%%%%%%%%%%%
\subsubsection{Residuals from Keplerian rotation}
%%%%%%%%%%%%%%%%%%%%%%%%%%%%%%%%%%%%%%%%%%%%%%
The best-fit \verb|discminer| models applied to each isotopolog (Table \,\ref{tab:params}) ultimately determined significantly varying disk inclinations and stellar masses among the individual tracers. These findings suggest that the disk may be warped, which we  discuss in more detail later in this work. Thus, to accurately compare the residual maps between the tracers, we re-run another two models for \twCO{} and \eiCO{} cubes, while fixing the geometry (offset, $i$, PA, $\upsilon_\mathrm{LSRK}$) to the best-fit \thCO{} parameters. This ensures that the resulting residuals between the tracers are not due to their different fitted geometric properties. A similar approach was applied in the exoALMA analysis \citep{izquierdo2025}, using the \twCO{} best-fit geometric parameters as reference. In our case, we regard the inferred \thCO{} geometry as more trustworthy compared to the perturbed \twCO{} kinematics and low S/N emission of \eiCO{}. Moreover, fitting the continuum visibility with a model that includes a ring and an asymmetry yielded an inclination and position angle consistent with the kinematic fit of \thCO{} \citep{Fasano_ea_2026}.

\begin{figure*}[t]
    \centering
    \includegraphics[width=1\linewidth]{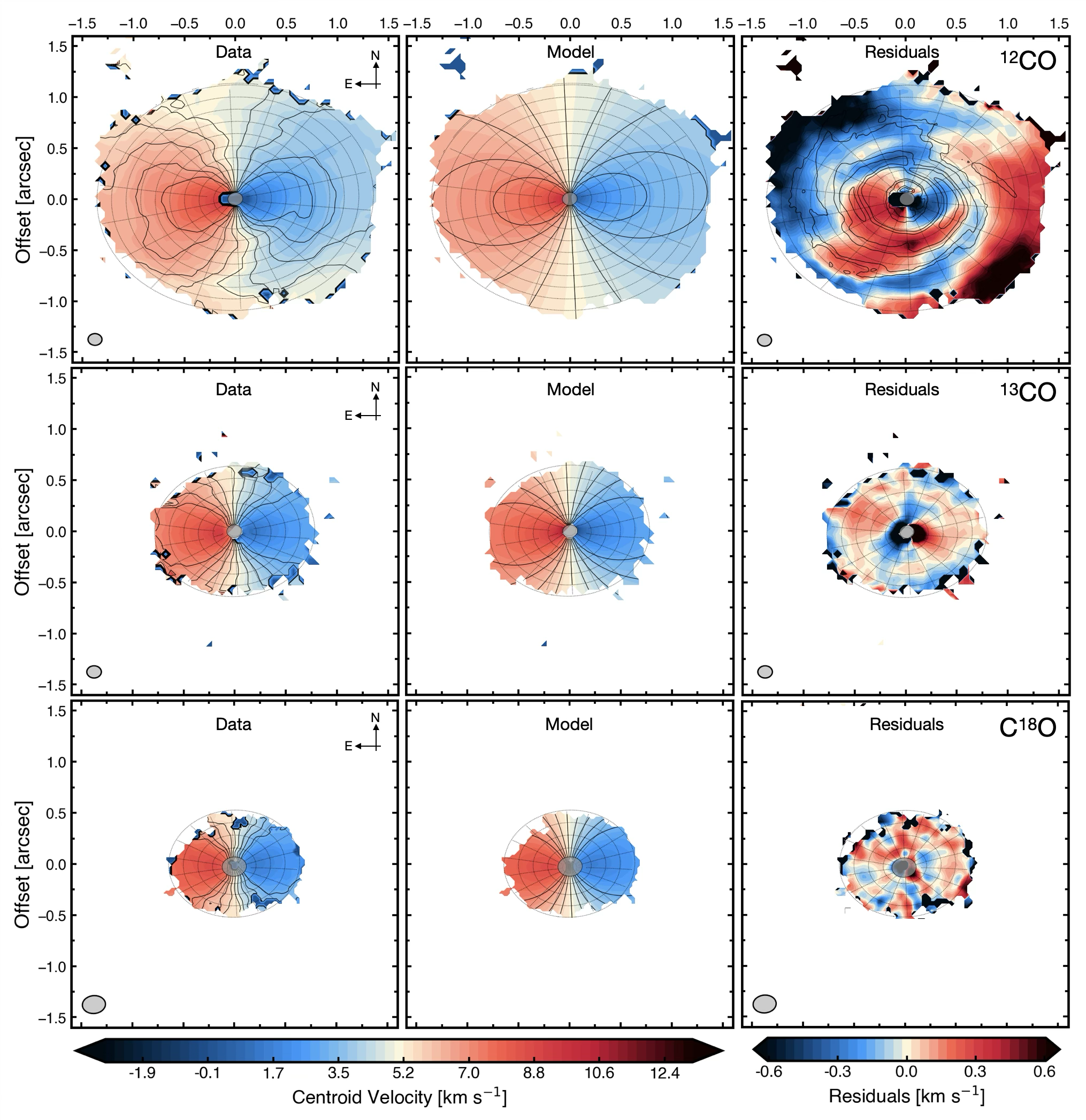}
    \caption{Gas kinematics for the CO molecular lines. Centroid LoS Gaussian moment maps for the data (left columns), Keplerian model (center columns), and data-model residuals (right columns), for the \twCO{} (top) and \thCO{} (middle) data, using high-resolution cubes. The geometry of \twCO{} and \thCO{} models was fixed to the \thCO{} disk, which is less perturbed and has a higher S/N than \eiCO{}. For \eiCO{} (bottom), we employ the fiducial cube. The respective beam sizes are shown in the lower left corners. In \twCO{}-residuals, we overplotted the IR contours from Fig.\,\ref{fig:overview}(a). All panels share the same scale, color bars, and are masked at 4$\sigma$ of their peak intensity map. The innermost beam size is masked in as it was excluded from the fit.}
    \label{fig:CO_residuals}  
\end{figure*}

Figure\,\ref{fig:CO_residuals} shows the Gaussian LoS centroid maps for the data (left panels), model (center panels), and residuals from Keplerian rotation (right panels) for all three isotopologs. The previous sections indicated that the \twCO{} residuals exhibit highly non-Keplerian motions. Indeed, spiral spurs are visible, outlining the IR ring (shown in gray contours). We also observed strong non-Keplerian residuals in \twCO{} at the location of the secondary line components to the southeast and a prominent red- (blue-) shifted pattern at the disk edges to the southwest (northeast). This pattern is further discussed later in this work in relation to disk warping. We additionally present the \twCO{} centroid residuals, determined by the LoS velocity from the channel of peak intensity shown in the ``quadratic'' moment maps in Fig.\,\ref{fig:12co_quad_resid}.

%%%%%%%%%%%%%%%%%%%%%%%%%%%%%%%%%%%%%%%%
%%%%%%%%%%%%%%%%%%%%%%%%%%%%%%%%%%%%%%%%

%%%%%%%%%%%%%%%%%%%%%%%%%%%%%%%%%%%%%%%%
\begin{figure*}[t]
    \centering
    \includegraphics[width=0.75\linewidth]{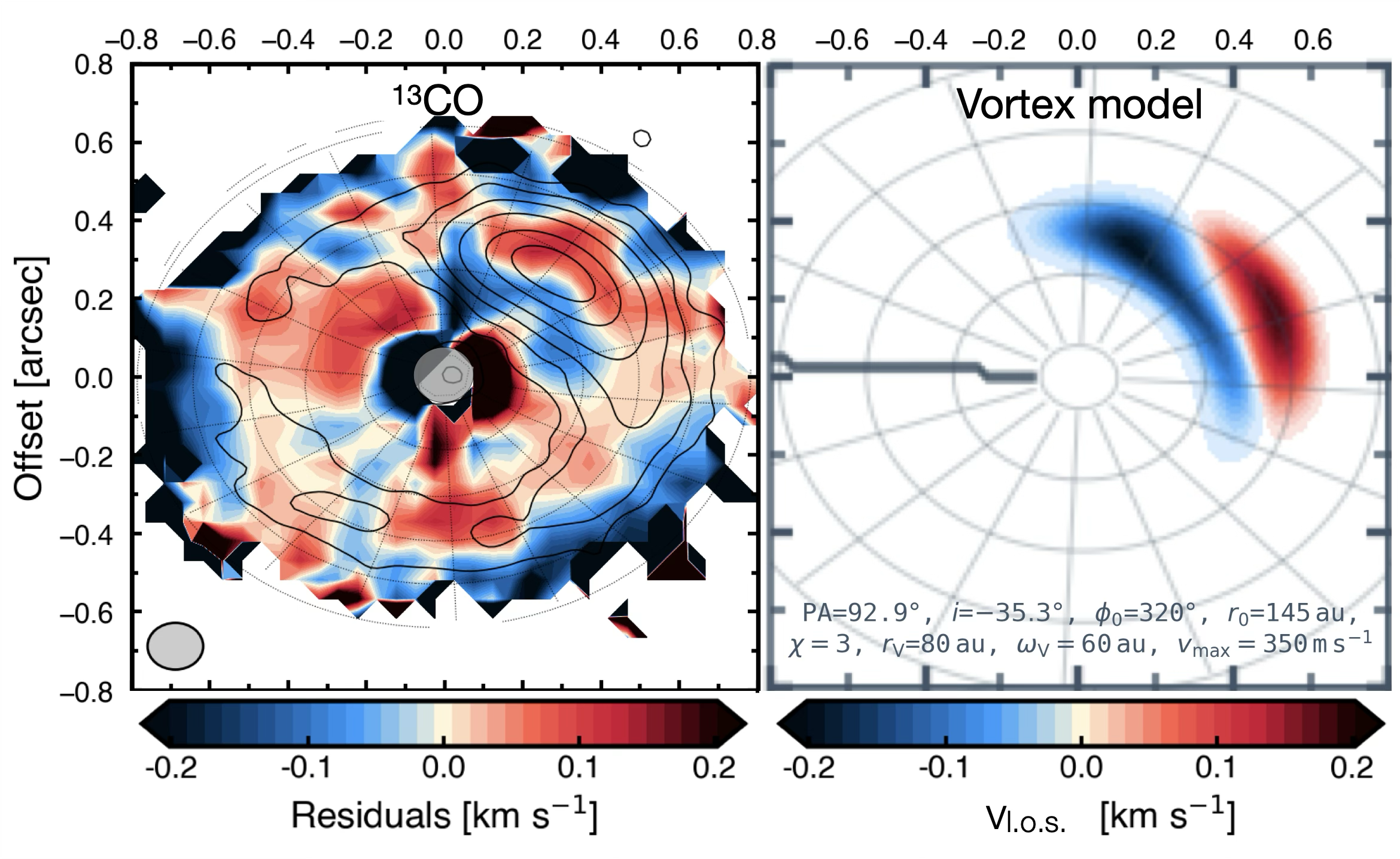}
    \caption{Vortex kinematics. Left: Centroid residuals for \thCO{} with overplotted continuum contours the same as in Fig.\,\ref{fig:overview}(c). Right: Line-of-sight velocities of the qualitatively best-matching vortex model \citep{Woelfer_ea_2025} with parameters at the bottom. Layout is the same as in Fig.\,\ref{fig:CO_residuals}.}
    \label{fig:vortex_kinematics}
\end{figure*} 
%%%%%%%%%%%%%%%%%%%%%%%%%%%%%%%%%%%%%%%%

The \thCO{} residuals show a blue or red lobe in the innermost regions, which can be attributed to sub-Keplerian rotation caused by the pressure gradient inside the gas cavity. Beyond a radius of approximately $R\approx 0\farcs5$, the residual pattern changes to reveal a tentative symmetric blue- or redshift ($\lesssim0.1\,$\kmsec{}) along the disk's minor axis, which might indicate a transition to super-Keplerian rotation coinciding with the IR and continuum ring. In contrast, the \eiCO{} residuals are difficult to interpret due to their patchy nature, presumably arising from this line's low S/N (see Table\,\ref{tab:imaging_params}).

%%%%%%%%%%%%%%%%%%%%%%%%%%%%%%%%%%%%%%%%%%%%%%
\subsubsection{Vortex kinematics}
%%%%%%%%%%%%%%%%%%%%%%%%%%%%%%%%%%%%%%%%%%%%%%
In this section, we focus on the kinematical patterns around the dust crescent. In Fig. \ref{fig:vortex_kinematics}, we zoom in on the velocity residuals of \thCO{}, emphasizing the dust continuum emission with overplotted contours. The red-blue velocity residual pattern co-located with the dust crescent has a magnitude of $\sim\pm0.10\,$\kmsec{}. Following Appendix B in \cite{Teague_ea_2022b}, we can estimate the significance of these residual velocities based on how accurately we can determine the line centroid from our observations. We measured a Gaussian width of $\Delta V=0.8\,$\kmsec{} and S/N$\approx12$ for the \thCO{} line co-located with the dust crescent. This allows us to determine the accuracy of our line centroid fitting to about one-fifth of a channel, or $\delta \upsilon_0 \approx 0.04\,$\kmsec{} \citep[see Fig.\,13 in][]{Teague_ea_2022b}. As a result, the observed residuals at the crescent are not significant, with a significance level of only \(\sigma = 0.10 / 0.04 \approx 2.5\). To improve on this result, observations at higher sensitivity and spectral resolution are needed to better sample the velocities at the crescent's location.

Still, we want to assess if this tentative residual pattern is suggestive of anticyclonic motions. Hence, we set up an analytical vortex model presented in \cite{Woelfer_ea_2025}, similar to the one developed by \cite{robert_ea_2020} and \cite{Boehler_ea_2021}, to assess whether the \thCO{} velocity residuals pattern is indicative of an anticyclonic vortex. This model describes the vortex with elliptic streamlines of constant velocity and a Gaussian velocity profile. It incorporates input parameters such as the radial, $r_\mathrm{0}$, and azimuthal, $\phi_\mathrm{0}$, locations of the vortex center, its Gaussian width, $\omega_v$, its aspect ratio $\chi$ (for $\chi=1$ it is circular), and the radius, $r_v$, of the ring of maximum velocity $\upsilon_\mathrm{max}$, as well as the disk geometry. We anchor the geometrical values following the dust continuum emission morphology.

We present the resulting $\upsilon_\mathrm{l.o.s.}$ vortex velocities, based on the location of the continuum crescent ($r_0=162\,$au, $\phi_0=-40\degree$) and a set of varying vortex parameters (see Fig.\,\ref{fig:vortex_models}). We attempted to fit the vortex model to the \thCO{} velocity residuals; however, it did not converge. Therefore, we present the model that best qualitatively (i.e., by-eye) aligns with the observed \thCO{} velocity residuals, as shown in the right panel of Fig. \ref{fig:vortex_kinematics}. Its vortex model parameters are $\omega_v=60\,$au, $\chi=3$, $r_v=100\,$au and $\upsilon_\mathrm{max}=350\,$\msec{}. We do not compare the \twCO{} residuals in the figure, as they are contaminated by the spiral features along the IR and continuum ring. We also disregard the \eiCO{} residuals due to their low sensitivity (S/N$<11$) and limited angular resolution. The \thCO{} velocity residual patterns are broadly consistent with those observed in the vortex model. This makes \sys{} an up-and-coming candidate for follow-up observations in deeper layers of the disk, which are less disturbed than its surface, allowing for a better study of vortex kinematics. Compared to two other major dust asymmetries, it is not affected by cloud absorption, unlike IRS\,48 \citep{vanderMarel_ea_2013}, and its dust crescent is located more favorably between the disk's minor and major axes to study radial and azimuthal motions when compared to HD\,142527 \citep{Boehler_ea_2021}. Finally, the CO molecular lines in \sys{} do not experience strong continuum extinction at the crescent location, unlike the other two systems.

%%%%%%%%%%%%%%%%%%%%%%%%%%%%%%%%%%%%%%%%%%%%%%
\section{Discussion}
\label{sec:discussion}
%%%%%%%%%%%%%%%%%%%%%%%%%%%%%%%%%%%%%%%%%%%%%%
The circumbinary disk \sys{} shows multiple spiral arms, signs of asymmetric continuum emission, and a large inner dust cavity. In the following discussion, we interpret our observations to reveal several key findings: 1) a highly asymmetric continuum crescent that an anticyclonic vortex may explain; 2) a gas cavity indicated by CO molecular lines; 3) \twCO{} line emission and kinematics that align with the scattered light spiral structures; and 4) whose line spectra show components of spirals and infalling material onto the upper layers of the disk's atmosphere. 

Previous studies on this system have attempted to understand the nature of the IR spirals of the circumbinary disk, in particular, their different and large opening angles, as well as the apparent ``break'' in the asymmetric IR ring toward the north \citep{monnier2019, uyama2020}. \cite{monnier2019} proposed the presence of a massive companion (50\,$M_\mathrm{J}$) at this location. However, such a companion has not been detected \citep{Columba_ea_2024}, and we do not observe any indications of it from our gas kinematics observations at this location, due to the absence of strong, clustered, high-variance deviations from Keplerian rotation \citep[e.g.,][]{Izquierdo_ea_2021}. We suggest that the break in the IR ring results from a geometrical viewing effect. The IR ring consists of several spiral arms that extend out of the plane, as observed in the H$\alpha$ image \citep{Columba_ea_2024}. Because the northern side of the disk is tilted toward us \citep{monnier2019}, it appears as if the two ends of the spirals are connected at the same radius, even though they are not likely to be in the same vertical plane (see Fig.\,\ref{fig:highres_dust_images}\,d).

We propose that the spirals and the ring observed in IR emission may indicate the presence of infalling material in the upper layers of the disk. Co-located with the IR ring, we can identify shoulders in the \twCO{} line spectra (shifted by a few \kmsec{}), as highlighted by the high reduced $\chi^2$ values in Fig. \ref{fig:line_profiles}. These shoulders are likely dominated by strong downward vertical motions of a few \kmsec{}, as indicated by pixels 2 and 3 being blueshifted and pixels 11 and 12 being redshifted. We consider the radial and azimuthal motions as physically unlikely, since when they are deprojected (divided by $\sin(i_0\approx35\degree)\approx0.6$), they would equate to motions in the disk plane of over 5\,\kmsec{}.

Recent hydrodynamical simulations, by \cite{Calcino_ea_2025}, have shown that such infall streamers drive long-lived ($\sim10^4$\,yr) super-Keplerian and stationary spiral arms of varying pattern speeds in the disks. \cite{Dullemond_ea_2019}  also suggested that cloudlet captures can create second-generation transition disks to explain their large cavities, which are generally otherwise hard to reproduce  with just one massive companion. Such disks would also be expected to be warped in their outer regions, since the angular momentum of the infalling material is misaligned with the initial disk structure. The significant differences in derived best-fit inclination and PA between each CO isotopolog (see Table \ref{tab:params}) might hint toward a warp in \sys{} that could help explain the prominent red and blue residual patterns in \twCO{} (Fig. \ref{fig:CO_residuals}), which appear beyond the IR contours and are nearly symmetrical around the disk's minor axis. A warp is characterized by changes in PA and inclination with radius. However, there are degeneracies between these geometric parameters and any real azimuthal or radial velocity perturbations that may still be present \citep{Rosenfeld_ea_2014}.

We present a fit to the \twCO{} velocity residuals using a simple warped disk model by \cite{Winter_ea_2025_warp} with radially varying inclination and position angle in Appendix \ref{app:warp_model}. We find a maximal nominal tilt of the disk of $\beta_\mathrm{max}=8.70\degree\pm0.27\degree (\pm1\sigma)$. As can be seen in Fig.\,\ref{fig:warp_model}, the warp model reproduces the outermost disk regions beyond 350\,au well, which exhibit a $m=1$ pattern in their centroid residuals. We expect the damping timescales of the warp's bending waves to be on the order of $\tau_\mathrm{damp}\sim1/\alpha\ \Omega_\mathrm{k}$ \citep{Ogilvie_Lubow_2002, Ogilvie_Latter_2013}. At $R=400\,$au and for $\alpha=10^{-3}$ around $M_\star=4\mathrm{M}_\odot$, this results in a relatively long-lived warp of $\tau_\mathrm{damp}\approx4$\,Myr. However, the more complex non-Keplerian features observed cannot be explained solely by the warp and are likely influenced by other dynamic processes, such as infall or a massive companion, potentially shaping this disk.

Our CO spectral analysis suggests the presence of additional non-disk-like (exodisk) velocity components, which we interpret as resulting from infalling gas streamers. Anisotropic infall has also been shown to trigger the RWI in disks and generate vortices in hydrodynamical simulations \citep{Bae2015, Kuznetsova2022}. Vortices are preferably formed near the radial locations of the infall zone, where the accretion streams hit the disk. We cannot state exactly where the potential streamers in the southeast and northeast hit the disk. In the south, it likely merges with the disk at the localized super-Keplerian \twCO{} residual feature located in the southern part of the IR ring (see Fig.\,\ref{fig:CO_residuals} and panels 9 and 10 Fig.\,\ref{fig:line_profiles}). We are uncertain about where the structure detected in the northeast connects with the disk, specifically whether it is at the front or the back. This structure might be a remnant of the infall. Further observations of shock-tracing molecules such as SiS and SiO, commonly employed in Class O/I sources \citep[e.g.,][]{Podio_ea_2017, Oya_ea_2019, Tychoniec_ea_2021}, or SO can help pinpoint these merging zones \citep{vanGelder_ea_2021}, as recently employed in the disk of AB\,Aur \citep{Speedie_ea_2025}. Interestingly, both transition disks share many similarities, including a cavity in the IR from which several spiral arms emanate, signs of infalling material, and a continuum ring with a strong asymmetry \citep{Tang2012, Tang2017, Calcino_ea_2025b}. The continuum crescent of \sys{} also closely resembles that of IRS\,48 \citep{vanderMarel_ea_2013, yang_ea_2023} and HD\,142527 \citep{boehler_ea_2017}, the most asymmetric continuum disks observed to date. HD\,142527 has a detected M dwarf companion \citep[][]{Lacour_ea_2016, Nowack_ea_2024} and theoretical works suggest the presence of a massive close-in companion in IRS\,48 to reproduce its observed features \citep{vanderMarel_ea_2013, calcino2019}. These parallels suggest that similar mechanisms might be at work shaping all three of these disks.

Another discussed scenario of the development of the spiral features is the flyby of one of \sys{}'s stellar companions. Three-dimensional hydrodynamic simulations indicate that flybys can trigger the formation of spiral arms that last for several dynamical timescales, which limits the temporal window for a close encounter to a several tens of thousands of years at most \citep{Smallwood_ea_2023, Cuello_ea_2019, Cuello_ea_2023}. Flybys primarily induce $m=2$ spiral arms \citep[e.g.,][]{Kurtovic_ea_2018} and both their pattern speed and pitch angles decay quickly over time. By analyzing Gaia DR3 proper motion measurements, \cite{shuai2022} found a stellar flyby within the past $10^{4}$\,yr for \sys{} unlikely. The closest approach ($d\approx3000\,{\rm au} \approx9\farcs0 $) was that of HD\,34700C (not B); however, we did not detect any disk around it. Furthermore, the most extended spiral arms observed in the IR display very large opening angles \citep[$\sim$30\degree{}][]{Columba_ea_2024}, which should have already wound up, given an optimistic flyby scenario \citep{Smallwood_ea_2023}. Lastly, theoretical studies suggest that flybys can cause disks to tilt and warp, yet the timescale of misalignments is too short-lived to explain the observations \citep{Nealon_ea_2020}.

Furthermore, the question remains of what causes the innermost gas cavity. Its edge can be placed inside the H$\alpha$ ring at $0\farcs20$ (65\,au), which aligns with the CO line emission drop, although the short-separation binary is not able to carve out such a large cavity \citep{Penzlin_ea_2024}. \cite{monnier2019}  suggested the possibility of another companion, located between the outer IR ring and the gas cavity, with current detection limits of a few Jupiter masses \citep{uyama2020, Columba_ea_2024}. Moreover, more massive companions within the gas cavity might be obscured by the coronagraph \citep[$r=92.5\,$mas$\approx32\,$au,][]{Columba_ea_2024}. These companions ought to be sub-stellar, with masses $\lesssim30\,$M$_\mathrm{J}$, with further evidence provided by the upper limits from the binary's astrometric proper motion anomaly \citep{Vioque_ea_2025}.

Our kinematic analysis shows strong residuals from Keplerian rotation  (Fig. \ref{fig:CO_residuals} and \ref{fig:vortex_kinematics}) interior of the continuum and IR ring ($R<0\farcs46\approx160\,$au) of the order of a few 100\,\msec{}. However, they are not straightforward to interpret; \twCO{} traces regions high up in the disk atmosphere, which is disturbed by the non-Keplerian motions of the spirals and the potentially infalling material. Nonetheless, the Keplerian deviations, seen as spiral spurs in the channel maps, are substantial along and inside the IR ring ($\sim2$\,\kmsec{}, Fig.\,\ref{fig:PV_diagram}). They display several abrupt changes in Keplerian rotation (blue- and redshifted patterns), called Doppler-flips, which potentially indicate the location of massive companions \citep{Perez_ea_2018}. However, subtracting a circular Keplerian model from highly non-Keplerian motions observed in an asymmetric, potentially warped, and eccentric disk can result in artificial velocity residuals and Doppler flips \citep{Norfolk2022, ragusa2024}. 

%%%%%%%%%%%%%%%%%%%%%%%%%%%%%%%%%%%%%%%%%%%%%%
\section{Conclusions}
\label{sec:conclusion}
%%%%%%%%%%%%%%%%%%%%%%%%%%%%%%%%%%%%%%%%%%%%%%
In this paper, we present the first high-resolution ALMA Band 6 observations of the dust continuum and the \twCO{}, \thCO{}, and \eiCOfull{} line emission of the circumbinary disk around \sys{}. Our key findings and conclusions are as follows.

\begin{enumerate}[i)]
    \item The dust continuum observations reveal a highly asymmetric dust crescent on top of a ring at $0\farcs39$ (138\,au) located at the edge of a cavity identified by IR observations \citep{monnier2019, Columba_ea_2024}. We further detected an unresolved inner disk co-located with the central binary at an 11$\,\sigma$ significance level. 
    
    \item The peak intensity maps of all CO isotopologs display an inner gas cavity of $R\approx65\,$au. The \twCO{} emission is highly structured and asymmetric, with the disk's southern part being brighter than its northern part. Its peak intensity exhibits a detached arc-shaped feature in the northeast and emission substructures that follow the spiral structures observed in the IR. 

    \item The \twCO{} kinematics shows strong non-Keplerian flows in the form of spiral spurs, which closely follow the IR ring, along with pronounced red- and blueshifted residuals at the disk's southwestern and northeastern edges. The emission morphology of the \thCO{} and \eiCO{} channel maps indicates that emissions from these isotopologs are less dynamically perturbed.

    \item We found varying disk position angles ($\Delta\rm{PA}_{18-12}\approx3$\degree{}) and inclinations ($\Delta i_{18-12}\approx10$\degree{}) inferred from the best-fit Keplerian models between \eiCO{} and \twCO{}. This could indicate that the disk is warped, further supported by our warp model fit \citep{Winter_ea_2025_warp} to the \twCO{} velocity residuals, which gives a maximum nominal tilt of the disk of $\beta_\mathrm{max}=8.7^{\circ}\pm0.3^{\circ}$.
    
    \item At the position of the continuum crescent, the \thCO{} velocity residuals show tentative signs of an anticyclonic vortex when compared to an analytical model \citep{Woelfer_ea_2025}. Future molecular line observations of brighter midplane tracers, combined with detailed hydrodynamic simulations, will help in determining whether a vortex would indeed be responsible for these observed residuals.

    \item The line spectra analysis of \twCO{} reveals a streamer above the southeastern disk plane, seen as a secondary component to its line profile, with its peak offset by over 1\,\kmsec{}. This infalling material may have triggered the Rossby Wave instability (RWI) at the cavity's edge \citep{Bae2015, Kuznetsova2022}, creating the potential vortex observed in the gas kinematics. Although we cannot pinpoint where this streamer hits the disk at present, future observations of chemical shock tracers, such as SO, can aid in identifying these locations \citep[e.g.,][]{Speedie_ea_2025}. 
    
\end{enumerate}
 
\noindent The nature of the depleted gas cavity remains a challenge. Our kinematic observations have not definitively pinpointed whether companions are responsible for shaping this cavity. We also speculate that \sys{} might host a second-generation disk that captured a cloudlet \citep[e.g.,][]{kuffmeier2020}. Consequently, this material may not have had sufficient time to accrete onto the binary \citep[$\log(\dot M_{\rm acc}/M_\odot\,{\rm yr}^{-1})=-6.8$;][]{Wichittanakom_ea_2020}, leading to the formation of the gas cavity. Future ALMA observations of brighter midplane tracers at higher resolution could help elucidate the complex kinematics of the cavity and the existence of an anticyclonic vortex.

\begin{acknowledgements}
The authors would like to thank the anonymous referee for the constructive feedback that enhanced the quality of this work. We also thank Alex Ziampras, Aleksandra Kuznetsova, Clement Baruteau, Enrico Ragusa, John Ilee, and Nicolás Cuello for their helpful discussions. We thank Gabriele Columba for providing the H$\alpha$ fits file, and Christian Ginski for the DESTINYS fits file and helpful comments. This project has received funding from the European Research Council (ERC) under the European Union’s Horizon 2020 research and innovation programme (PROTOPLANETS, grant agreement No. 101002188). JS has performed computations on the `Mesocentre SIGAMM' machine, hosted by Observatoire de la Côte d’Azur, and acknowledges assistance from Allegro, the European ALMA Regional Center node in the Netherlands.
AJW has been supported by the Royal Society through a University Research Fellowship grant number URF\textbackslash R1\textbackslash 241791.
Support for AFI was provided by NASA through the NASA Hubble Fellowship grant No. HST-HF2-51532.001-A awarded by the Space Telescope Science Institute, which is operated by the Association of Universities for Research in Astronomy, Inc., for NASA, under contract NAS5-26555.
JeS acknowledges financial support from the Natural Sciences and Engineering Research Council of Canada (NSERC) through the Canada Graduate Scholarships Doctoral (CGS D) program, and from the Heising–Simons Foundation through the 51 Pegasi b Fellowship.
JB acknowledges support from NASA XRP grant No. 80NSSC23K1312.
SF is funded by the European Union (ERC, UNVEIL, 101076613). SF acknowledges financial contribution from PRIN-MUR 2022YP5ACE.
This paper makes use of the following ALMA data: ADS/JAO.ALMA\#2022.1.00760.S. ALMA is a partnership of ESO (representing its member states), NSF (USA), and NINS (Japan), together with NRC (Canada),  NSC and ASIAA (Taiwan), and KASI (Republic of Korea), in cooperation with the Republic of Chile. The Joint ALMA Observatory is operated by ESO, AUI/NRAO, and NAOJ. \\

\textit{Software}: \texttt{analysisUtils} \citep{Hunter_ea_2023}, \texttt{Bettermoments} \citep{bettermoments}, \texttt{CARTA} \citep{Comrie_2021_carta}, \texttt{CASA} \citep{McMullin2007, CASA_Team_ea_2022}, \texttt{Discminer} \citep{Izquierdo_ea_2021}, \texttt{emcee} \citep{emcee2013, Emcee_2019}, \texttt{GoFish} \citep{Teague_2019_gofish}, \texttt{Matplotlib} \citep{Hunter_mpl}, \texttt{Numpy} \citep{vanderWalt_np}, \texttt{reduction\_utils} scripts by the DSHARP and MAPS ALMA Large Programs \citep{Andrews_ea_2018, czekala2021}, \texttt{Scipy} \citep{Virtanen_scipy}.
\end{acknowledgements}

%\bibpunct{(}{)}{;}{a}{}{,} % to follow the A&A style
\bibliographystyle{aa} % style aa.bst
\bibliography{mybib} % your

%%%%%%%%%%%%%%%%%%%%%%%%%%%%%%%%%%%%%%%%%%%%%%%%%%%%%%%%%%%%%%%%%%%%%%%%%%%%%%%%%%%%%%%%%%%%
\begin{appendix} %First Appendix
\onecolumn

%%%%%%%%%%%%%%%%%%%%%%%%%%%%%%%%%%%%%%%%%%%%%%
\section{Observations, imaging, and emission properties}
\label{app:obs}
%%%%%%%%%%%%%%%%%%%%%%%%%%%%%%%%%%%%%%%%%%%%%%

Table~\ref{tab:observations} contains the relevant information of each EB used for the measurement sets and images produced. Table~\ref{tab:imaging_params} lists the imaging parameters and emission properties of the continuum and molecular line data cubes for both HD\,34700A and B.

%%%%%%%%%%%%%%%%%%
\begin{table*}[h]
\centering
\begin{threeparttable}
{\renewcommand{\arraystretch}{1.3}%pad between rows
\caption{Details of all the EBs included in the HD\,34700 measurement set.} \label{tab:observations}
\begin{tabular}{ccccccccc}
\hline \hline
 Date &  No. Ant. &  Int &  Baselines & Resolution & Max. Scale &  Phase Cal. &  Flux/Bandpass Cal.  \\
  &   &  \footnotesize (min) &  \footnotesize (m) &  \footnotesize (arcsec) &  \footnotesize (arcsec) &  &  
\\
\hline
2022-10-09 & 43 & 50  & 15-500  & 0.64 & 6.97 & {\rm J}0509+0541 & {\rm J}0423-0120 \\
2022-10-09 & 43 & 50  & 15-500  & 0.64 & 6.97 & {\rm J}0527+0331 & {\rm J}0423-0120 \\

2023-05-13 & 42 & 47  & 15-2516  & 0.13 & 2.12 & {\rm J}0509+0541 & {\rm J}0423-0120 \\
2023-05-19 & 44 & 47  & 78-3638  & 0.10 & 1.43 & {\rm J}0509+0541 & {\rm J}0423-0120 \\
2023-05-19 & 44 & 47  & 78-3638  & 0.10 & 1.43 & {\rm J}0509+0541 & {\rm J}0423-0120 \\
2023-05-19 & 44 & 47  & 78-3638  & 0.10 & 1.43 & {\rm J}0509+0541 & {\rm J}0423-0120 \\
2023-05-21 & 42 & 47  & 28-3638  & 0.10 & 1.44 & {\rm J}0509+0541 & {\rm J}0423-0120 \\
2023-05-21 & 42 & 47  & 28-3638  & 0.10 & 1.44 & {\rm J}0509+0541 & {\rm J}0423-0120 \\
2023-05-24 & 42 & 47  & 28-3638  & 0.09 & 1.41 & {\rm J}0509+0541 & {\rm J}0423-0120 \\
2023-05-28 & 42 & 47  & 28-3638  & 0.10 & 1.61 & {\rm J}0509+0541 & {\rm J}0423-0120 \\
2023-05-28 & 42 & 47  & 28-3638  & 0.10 & 1.61 & {\rm J}0509+0541 & {\rm J}0423-0120 \\
2023-05-30 & 42 & 47  & 28-3638  & 0.10 & 1.44 & {\rm J}0509+0541 & {\rm J}0423-0120 \\
2023-06-02 & 42 & 47  & 28-3638  & 0.10 & 1.57 & {\rm J}0509+0541 & {\rm J}0423-0120 \\

\hline
\end{tabular}
}
\end{threeparttable}
\end{table*}
%%%%%%%%%%%%%%%%%%

%%%%%%%%%%%%%%%%%%
\begin{table*}[h]
\centering
\begin{threeparttable}
{\renewcommand{\arraystretch}{1.3}%pad between rows
\caption{HD\,34700 imaging parameters and emission properties of various tracers.} \label{tab:imaging_params}
\begin{tabular}{c c c c c c c c c c c}
\hline \hline
\multirow{2}{*}{Source} & \multirow{2}{*}{Tracer} & \multicolumn{1}{c}{Freq.} & \multicolumn{1}{c}{Chan. Sp.} & \multirow{2}{*}{Rob.} & \multicolumn{1}{c}{Beam size} & \multicolumn{1}{c}{PA} & \multicolumn{1}{c}{RMS Noise} & \multicolumn{1}{c}{Peak int.} & \multirow{2}{*}{S/N} & \multicolumn{1}{c}{Int. Flux} \\
&  & \multicolumn{1}{c}{(GHz)} & \multicolumn{1}{c}{(km/s)} &  & \multicolumn{1}{c}{(mas$^2$)} & \multicolumn{1}{c}{(deg)} & \multicolumn{1}{c}{(mJy\,bm$^{-1}$)} & \multicolumn{1}{c}{(mJy\,bm$^{-1}$)} &  & \multicolumn{1}{c}{(mJy)}\\
(1) & (2) & (3) & (4) & (5) & (6) & (7) & (8) & (9) & (10) & (11) \\
\hline 
AaAb & Continuum & 225.3 & 1.5 & 0.5 & $137 \times 115$ & 86.3 & $5.07 \times 10^{-3}$ & 2.32 & 458 & 8.42$\pm0.84$ \\
 & \twCO{} & 230.5 & 0.1 & 0.5 & $137 \times 115$ & 88.8    & 1.12 & 29.9 & 26.7 &  $4167\pm42$ \\
  & \thCO{} & 220.4 & 0.2 & 0.5 & $142 \times 119$ &  90.1  & 0.85 & 18.8 & 22.1 & $805\pm80$ \\
   & \eiCO{} & 219.6 & 0.2 & 1.5 & $227 \times 175$ &  95.2 & 0.68 & 16.9 & 10.9 & $123\pm12$ \\
\hline
B & Continuum & 225.3 & 1.5 & 0.5 & $137 \times 115$ & 86.3 & $5.07 \times 10^{-3}$ & 1.61 & 318 & $3.12\pm0.31$ \\
& \twCO{} & 230.5 & 0.1 & 0.5 & $137 \times 115$ &  88.8  & 1.12 & 13.1 & 11.7 & $106\pm11$  \\
\hline
\end{tabular}
}
\begin{tablenotes}
      \small
      \item Columns - (1): Source. (2): Tracer. (3): Average observation frequency. (4): Imaged channel spacing. (5): Briggs weighting robust parameter. (6) and (7): Synthesized beam size and PA. (8): RMS noise. (9): Peak intensity. (10): Peak S/N. (11): Integrated flux with 10\% flux uncertainty in ALMA Band 6.
    \end{tablenotes}
\end{threeparttable}
\end{table*}
%%%%%%%%%%%%%%%%%%

%%%%%%%%%%%%%%%%%%%%%%%%%%%%%%%%%%%%%%%%%%%%%%
\section{Circumstellar disk of HD 34700B}
\label{app:hd34700b}
\begin{figure*}[h!]
    \centering
    \includegraphics[width=0.70\linewidth]{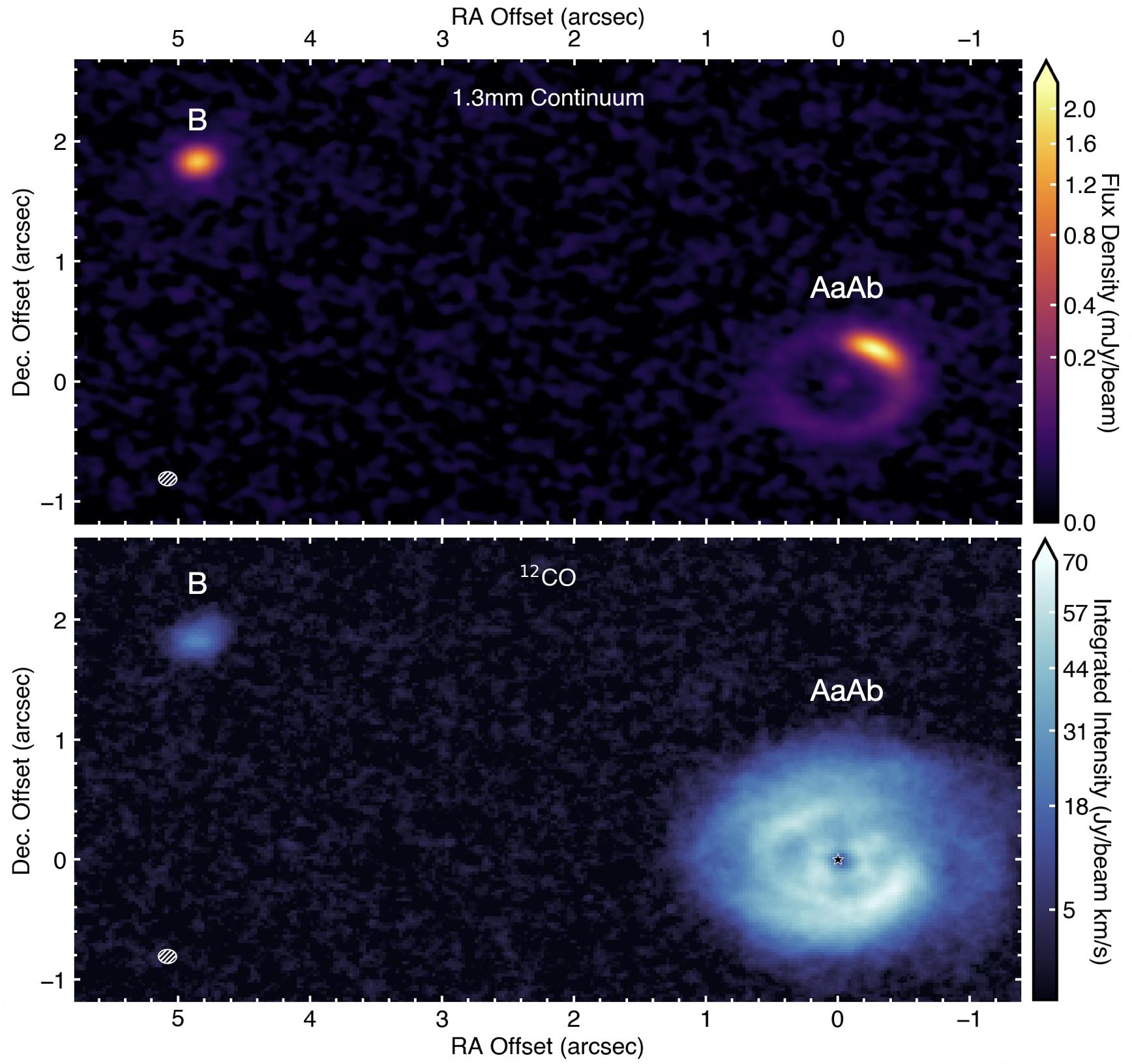}
    \caption{Overview plot of HD\,34700 AaAb and B. Top: 225.3\,GHz dust continuum emission plotted with a power-law scaling of $\gamma=0.4$ to highlight faint emission. Bottom: \twCOfull{} moment 0 map. The beam size ($0\farcs14\times0\farcs11$) of the robust 0.5 cubes employed is shown in the lower-left corner. No continuum or \twCO{} molecular line emission was detected bridging both disks around HD\,34700A and B, even when the data were imaged at a lower resolution and higher sensitivity.}
    \label{fig:HD34700AB_largescale}
\end{figure*}

\begin{figure*}[h!]
    \centering
    \includegraphics[width=1.0\linewidth]{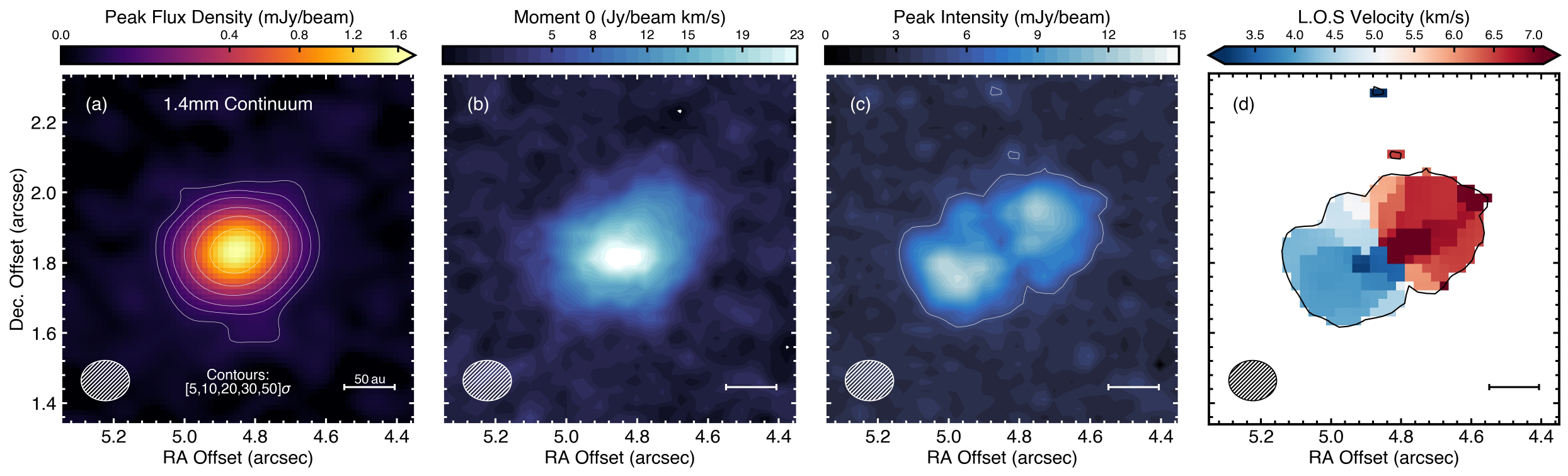}
    \caption{Band 6 observation of HD\,34700B. \textbf{(a)} 225.3\,GHz dust continuum. \textbf{(b)} \twCO{} moment 0 map. \textbf{(c)} \twCO{} peak intensity map, white contour encloses 5\,$\sigma$. \textbf{(d)} \twCO{} LoS quadratic centroid map, masked at 5$\sigma$ of the peak intensity map. The beam size of the high-resolution cubes employed is shown in the lower left corner.}
    \label{fig:HD34700B_allmom}
\end{figure*}

%%%%%%%%%%%%%%%%%%%%%%%%%%%%%%%%%%%%%%%%%%%%%%
We present an overview plot of the HD\,34700 system showing \sys{} and B in the same field of view in Fig.\,\ref{fig:HD34700AB_largescale}. We re-detected both continuum and \twCO{} emission co-located with companion B, its peak intensities and integrated fluxes are listed in Table\,\ref{tab:imaging_params}. We did not detect any \thCO{} or \eiCO{} emission ($>3\sigma$) around B, and no CO molecular line or continuum emission around HD\,34700C. In addition, we did not detect any continuum or \twCO{} emission bridging over A to B, even when the data were imaged at lower resolution and higher sensitivity. In Fig.\,\ref{fig:HD34700B_allmom}, we show the Band 6 detected continuum and \twCO{} observations centered and zoomed-in on HD\,34700B. The LoS velocity (Fig.\,\ref{fig:HD34700B_allmom}d) shows the clockwise Keplerian rotation of the disk with a systemic velocity $\upsilon_\mathrm{LSRK}$ similar to that of \sys{} (colorbar centered on $\upsilon_\mathrm{LSRK}$ of A), confirming that the stars are co-moving. Due to the low S/N and resolution, we did not attempt to fit for \twCO{} emission around B.

%%%%%%%%%%%%%%%%%%%%%%%%%%%%%%%%%%%%%%%%
%%%%%%%%%%%%%%%%%%%%%%%%%%%%%%%%%%%%%%%%

%%%%%%%%%%%%%%%%%%%%%%%%%%%%%%%%%%%%%%%%
%%%%%%%%%%%%%%%%%%%%%%%%%%%%%%%%%%%%%%%%

%%%%%%%%%%%%%%%%%%%%%%%%%%%%%%%%%%%%%%%%%%%%%%
\section{Best-fit parameter table}
\label{app:best-fit params}
%%%%%%%%%%%%%%%%%%%%%%%%%%%%%%%%%%%%%%%%%%%%%%
In Table\,\ref{tab:params} we show the best-fit \verb|discminer| model parameters of \sys{} for each CO isotopolog.

%%%%%%%%%%%%%%%%%%
\begin{table*}[h!]
\centering
\begin{threeparttable}
{\renewcommand{\arraystretch}{1.3}%pad between rows
\caption{Best-fit \texttt{discminer} model parameters.} \label{tab:params}
\begin{tabular}{|l|cc|rrrr|rrrr|}
\hline
\multirow{2}{*}{Tracer} & \multicolumn{1}{c}{$M_{\star}$} & \multicolumn{1}{c|}{$\upsilon_{\rm LSRK}$} & \multicolumn{1}{c}{$i_0$} & \multicolumn{1}{c}{PA} & \multicolumn{1}{c}{$x_{\rm c}$} & \multicolumn{1}{c|}{$y_{\rm c}$} & \multicolumn{1}{c}{$z_0$} & \multicolumn{1}{c}{$p$} & \multicolumn{1}{c}{$R_{\rm t}$} & \multicolumn{1}{c|}{$q$} \\
     & $({\rm M}_{\odot})$ & (km\,s$^{-1}$) & \multicolumn{1}{c}{(deg)} & \multicolumn{1}{c}{(deg)} & \multicolumn{1}{c}{(mas)} & \multicolumn{1}{c|}{(mas)} & \multicolumn{1}{c}{(au)} & & \multicolumn{1}{c}{(au)} & \\
\hline
\twCO{}    & 4.38$_{-0.02}^{+0.02\dagger}$ & 5.125$^{+0.002}_{-0.001}$ & $-30.9^{+0.1}_{-0.1}$ & $96.5^{+0.1}_{-0.1}$  & $-26^{+4}_{-1}$ & 8$^{+1}_{-2}$ & 12.1$^{+0.3}_{-1.0}$ & 0.34$^{+0.04}_{-0.01}$ &  331$^{+4}_{-11}$ & 1.57$^{+0.01}_{-0.01}$  \\
\thCO{}    & 4.64$^{+0.04}_{-0.05}$ & 5.122$^{+0.002}_{-0.003}$ & $-35.3^{+0.2}_{-0.2}$ & $93.0^{+0.1}_{-0.1}$  & $-3.1^{+0.8}_{-0.7}$ & $30^{+2}_{-2}$ & $62^{+2}_{-3}$ & $0.86^{+0.03}_{-0.03}$ & $125^{+6}_{-5}$ & $0.95^{+0.02}_{-0.02}$ \\
\eiCO{}    & 3.92$_{-0.06}^{+0.08\dagger}$ & 5.121$^{+0.011}_{-0.008}$ & $-42.6^{+0.4}_{-0.7}$ & $94.0^{+0.2}_{-0.3}$  & $7^{+2}_{-2}$ & $47^{+2}_{-26}$ & $
67^{+4}_{-27}$ & $0.34^{+0.03}_{-0.02}$ & $153^{+14}_{-14}$  & $0.50^{+0.09}_{-0.05}$
\\
\hline
\end{tabular}
}
\begin{tablenotes}
      \small
      \item The total stellar mass of the binary $M_\star$, systemic velocity $\upsilon_{\rm LSRK}$, geometrical parameters ($x_{\rm c}, y_{\rm c}$ offset from the central binary, PA position angle measured from the North to the redshifted major axis, and negative inclination $i_0$ indicates anti-clockwise disk rotation), and emission surface parameters ($z_0,p, R_{\rm t},q$ describe an exponentially tapered power law) were inferred from the high-resolution cubes for \twCO{} and \thCO{} and the fiducial cube for \eiCO{}. The reported MCMC uncertainties are represented by the 16th and 84th percentiles drawn from the posterior distribution of the final 10\% walkers of each model. These \texttt{discminer}-model uncertainties can increase by a factor of $\sim10$ due to spatially correlated noise, as shown by \cite{Hilder_ea_2025}. The strongest correlation between model parameters is observed between the stellar mass and the disk inclination, which are known to be degenerate \citep[e.g.][]{Teague_ea_2018c}. This $M_\star\cdot \sin i_0$ degeneracy is evident through the varying best-fit parameters among CO isotopologs. Note that the inferred inclinations $i_0$ and position angles PA between the CO tracers are significantly different, which we interpret as a sign of disk warping.
      \item ${\dagger}$ Stellar masses for \twCO{} and \eiCO{} models with geometrical parameters fixed to the best-fit \thCO{} ones are 3.58\,$M_\odot$ and 5.01\,$M_\odot$, respectively. 
    \end{tablenotes}
\end{threeparttable}
\end{table*}
%%%%%%%%%%%%%%%%%%

%%%%%%%%%%%%%%%%%%%%%%%%%%%%%%%%%%%%%%%%%%%%%%
\section{Warped disk model}
\label{app:warp_model}
%%%%%%%%%%%%%%%%%%%%%%%%%%%%%%%%%%%%%%%%%%%%%%
\cite{Winter_ea_2025_warp} demonstrated that the $m=1$ LoS velocity residual structure, after subtracting a Keplerian model, can be interpreted as evidence of moderate disk warping. In this section of the appendix, we present an interpretation of the \twCO{} velocity residuals (Fig.\,\ref{fig:CO_residuals}) as due to warping of the disk, which means a moderate change of the disk's inclination $i_0$ and position angle PA with radius. We employ the analytical warp model recently published by \cite{Winter_ea_2025_warp} within the exoALMA Large Program paper series \citep{teague2025}. The model interprets velocity residual motions $\delta\upsilon_\mathrm{los}$ as primary due to a varying LoS component of the Keplerian rotational velocity $\upsilon_\phi(R)=\sqrt{GM_\star/R}$ due to disk warping. We employ the \twCO{} \verb|discminer| model with geometrical properties fixed to the \thCO{}, as it most robustly traces the underlying Keplerian disk rotation. Minor changes to the best-fit inclination $\delta i_0$ and position angle $\delta\mathrm{PA}$ will then tilt each radial annulus compared to the flat initial LoS velocity $\upsilon^\mathrm{flat}_\mathrm{los}(R,\phi)$. The residuals in the LoS velocity can then be described as

\begin{equation}
    \delta\upsilon_\mathrm{los} (R,\phi)=\upsilon^\mathrm{warp}_\mathrm{los} - \upsilon^\mathrm{flat}_\mathrm{los} =\upsilon_\phi(R)\left[\delta i(R) \cos i_0 \cos \phi+\delta \mathrm{PA}(R) \sin i_0 \sin \phi\right] .
\end{equation}

The angle $\phi$ is measured counter-clockwise from the redshifted major axis of the disk ($\phi=0$). The above equation can also be rewritten in the form 
\begin{equation}
\delta \upsilon_{\operatorname{los}}(R, \phi)=A(R) \cos \phi+B(R) \sin \phi .
\end{equation}

The warp model then fits the coefficients $A(R)$ and $B(R)$ to each radial annulus of the velocity residual map, which are given by

\begin{equation}
\begin{aligned}
\delta i(R) & =\frac{A(R)}{\upsilon_\phi(R) \cos i_0}, \\
\delta \mathrm{PA}(R) & =\frac{B(R)}{\upsilon_\phi(R) \sin i_0}.
\end{aligned}
\end{equation}

We can further define the warp's tilt angle $\beta(R) \approx \sqrt{\delta i(R)^2+\delta \mathrm{PA}(R)^2 \sin ^2 i_0}$ with its maximum value $\beta_\mathrm{max}$ interpreted as the tilt amplitude \citep{Winter_ea_2025_warp}.

\begin{figure*}[t]
    \centering
    \includegraphics[width=0.8\linewidth]{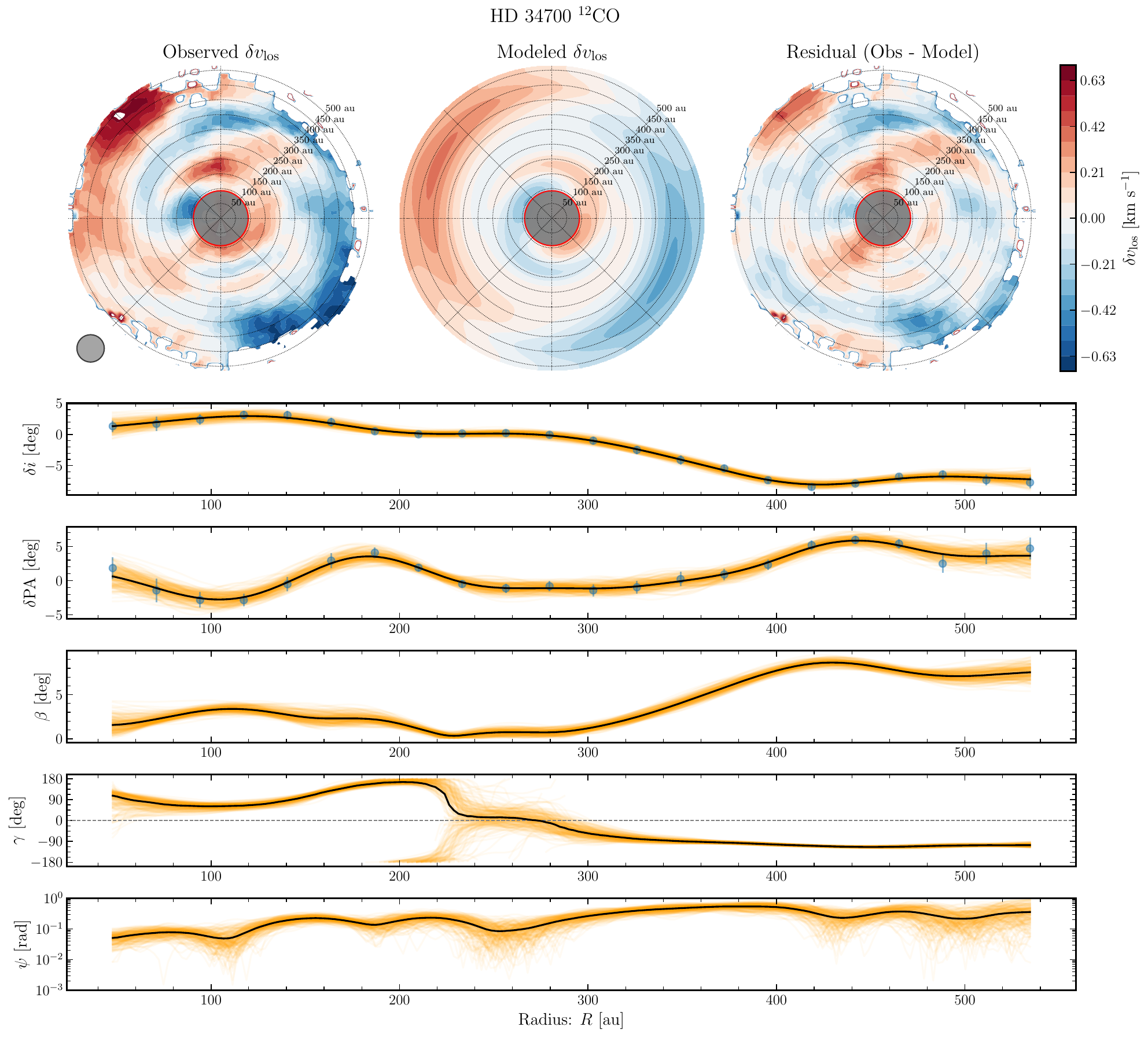}
    \caption{Warped disk model for \twCO{} HD~34700A. Top panels: Residuals (right) from the observed (left) vs modeled (center) $\delta \upsilon_{\rm los}$ fields for HD~34700A after fitting Keplerian velocity profiles. The color scale $\delta\upsilon_\mathrm{los}$ in km/s is identical in all panels. Grey circles mask twice the central beam size. The beam size ($0\farcs14\approx50\,$au) is also shown on the left-hand side, assumed to be circular for visualization. Bottom panels: Radial profiles of \( \delta i \), \( \delta \mathrm{PA} \), and the physical warp properties tilt $\beta$, twist $\gamma$, and the warp amplitude $\psi$ from our fitting procedure \citep[Eqs. 13-15 in][]{Winter_ea_2025_warp}. Blue points and error bars in \( \delta i \) and \( \delta \mathrm{PA} \) come from the least-squares fitting procedure. Faint orange lines show posterior distributions from a Gaussian process model.}
    \label{fig:warp_model}
\end{figure*}

The best-fit warp parameters and the resulting LoS residual map $\delta\upsilon_{\operatorname{los}}$ are shown in Fig.\,\ref{fig:warp_model}. The warp model residual map reproduces the strong $m=1$ residual pattern in the outermost disk regions beyond 350\,au well, represented by changes in disk inclination and position angle of up to 8\degree{} from their best-fit values. However, inside 300\,au, no clear $m=1$ patterns are observed, and thus the LoS residuals cannot be explained by warping alone. Overall, there are significant residuals between the observed data and the warped model. Other dominant dynamics, such as infall, likely drive these differences. Since the model attributes the entire LoS velocity difference ($\delta\upsilon_{\operatorname{los}}$) to warping, it is likely that the warp's tilt ($\beta$) and amplitude ($\psi$) are overestimated.

%%%%%%%%%%%%%%%%%%%%%%%%%%%%%%%%%%%%%%%%
%%%%%%%%%%%%%%%%%%%%%%%%%%%%%%%%%%%%%%%%

\newpage
%%%%%%%%%%%%%%%%%%%%%%%%%%%%%%%%%%%%%%%%%%%%%%
\section{Complementary figures}
\label{app:figures}
%%%%%%%%%%%%%%%%%%%%%%%%%%%%%%%%%%%%%%%%%%%%%%

%%%%%%%%%%%%%%%%%%%%%%%%%%%%%%%%%%%%%%%%
\begin{figure*}[h!]
    \centering
    \includegraphics[width=0.95\linewidth]{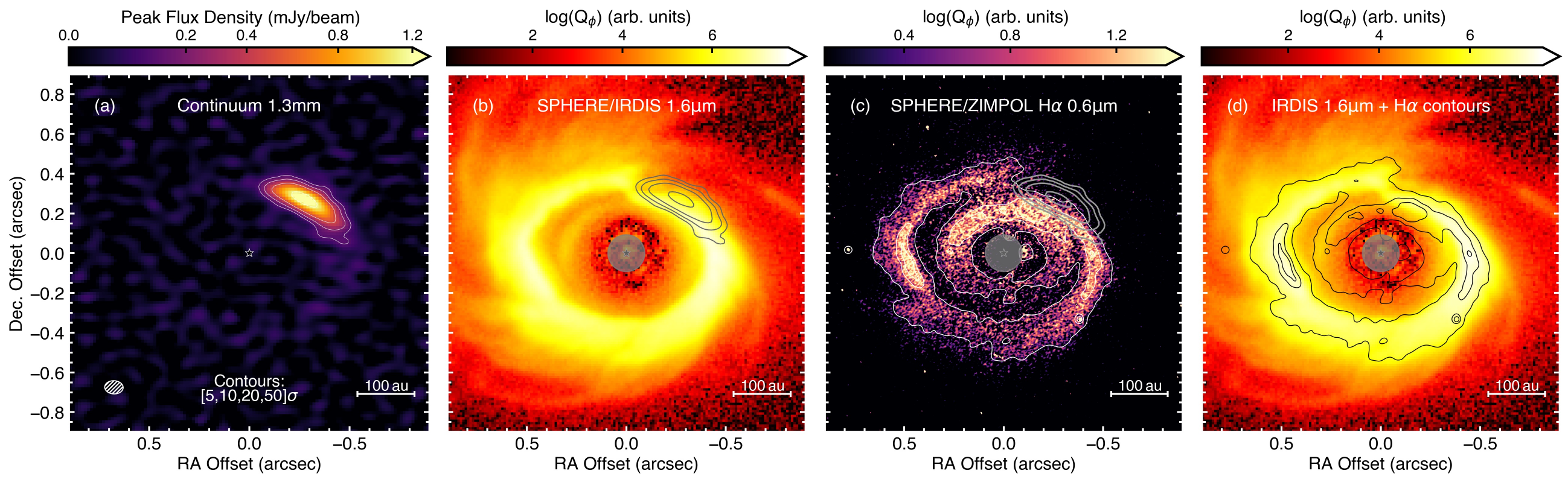}
    \caption{Gallery of \sys{} dust images. \textbf{(a)} High-resolution ($92\times66\,$mas$^{2}$) 225.3\,GHz dust continuum. \textbf{(b)} SPHERE/IRDIS IR scattered light observations and \textbf{(c)} SPHERE/ZIMPOL H$\alpha$ observations presented in \citet{Columba_ea_2024}, with the gray circles representing the coronograph ($r=92.5\,$mas\,$\approx32\,$au).  Overlaid dust continuum contours are the same as in (a). \textbf{(d)} IR scattered light observations with H$\alpha$ contours of (c). }
    \label{fig:highres_dust_images}
\end{figure*} 
%%%%%%%%%%%%%%%%%%%%%%%%%%%%%%%%%%%%%%%%

%%%%%%%%%%%%%%%%%%%%%%%%%%%%%%%%%%%%%%%%
\begin{figure*}[h!]
    \centering
    \includegraphics[width=0.95\linewidth]{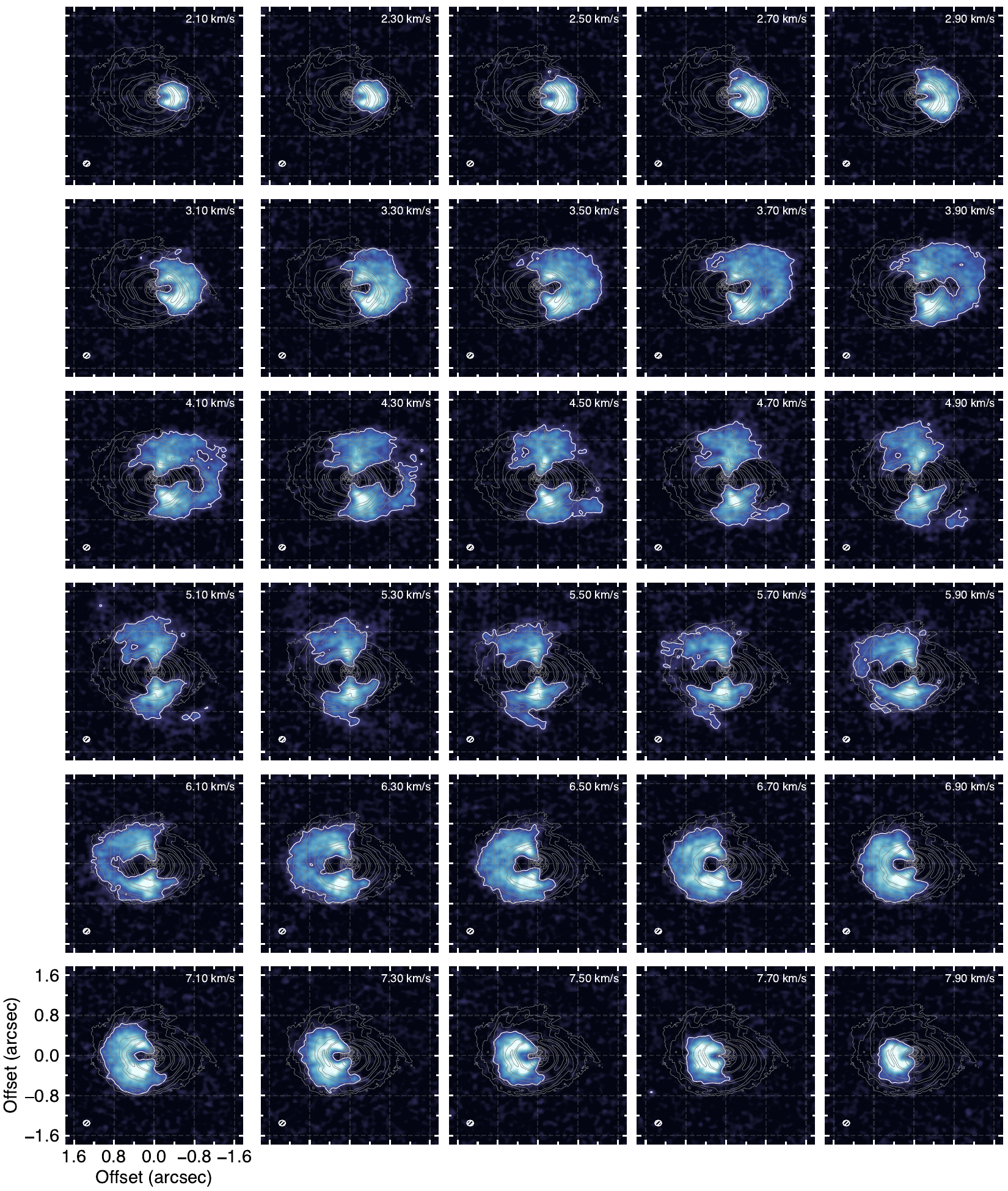}
    \caption{Channel maps for the \twCOfull{} molecular line. The white contour enclosing the line emission shows 5$\sigma$. The gray background contours indicate the scattered light emission, same as Fig.\,\ref{fig:overview}\,(a), to highlight that the spiral spurs in the line emission closely follow the IR ring. The beam size of the high-resolution cube is plotted in the lower left corner. Note that we only show every second channel (200\,\msec{}), and the systemic-velocity channel is at 5.1\,\kmsec{}. A movie of the \twCO{} channel maps can be found in the online resources.}
    \label{fig:12co_channels}
\end{figure*} 
%%%%%%%%%%%%%%%%%%%%%%%%%%%%%%%%%%%%%%%%

%%%%%%%%%%%%%%%%%%%%%%%%%%%%%%%%%%%%%%%%
\begin{figure*}[t]
    \centering
    \includegraphics[width=1.0\linewidth]{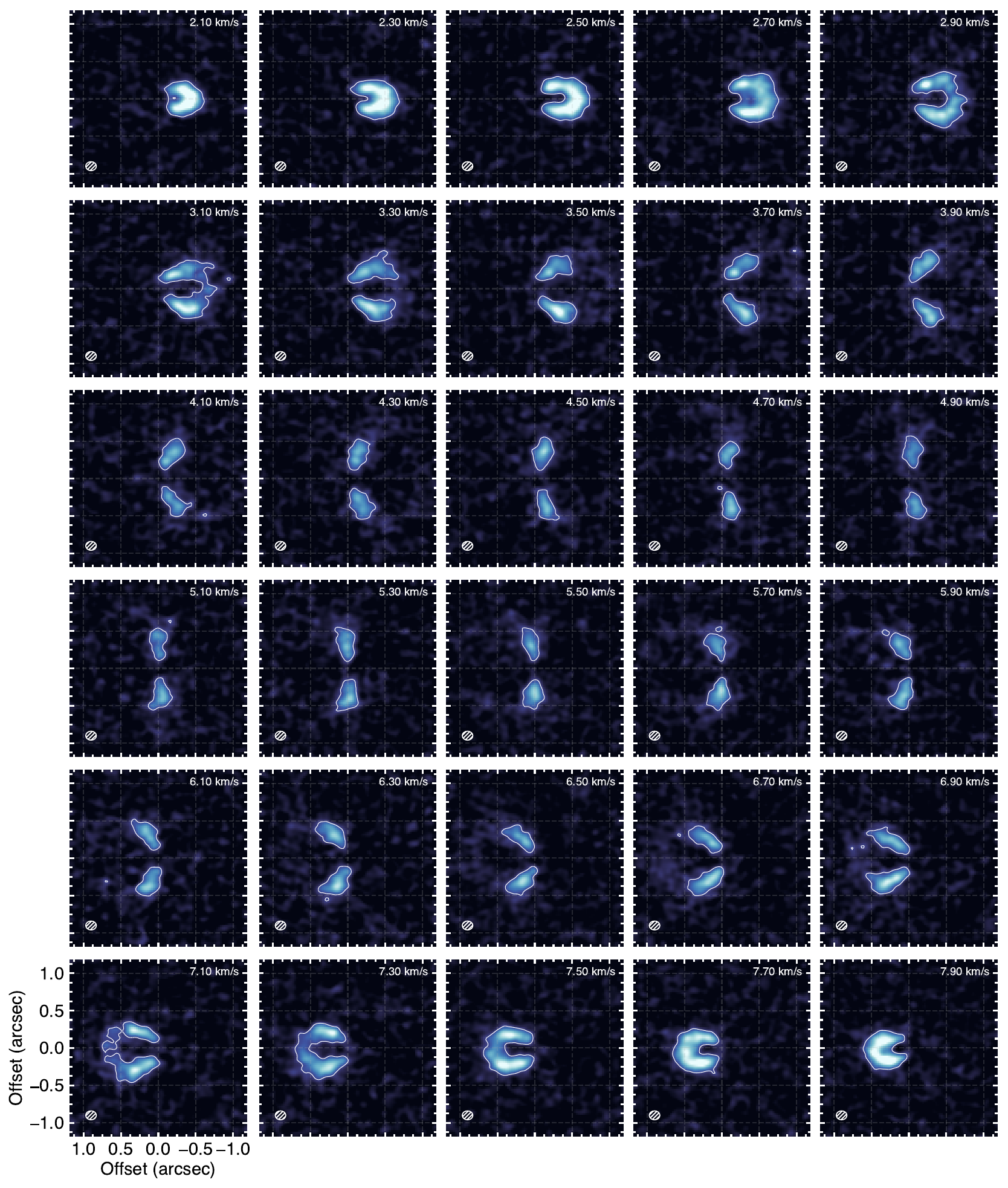}
    \caption{Same as Fig.\,\ref{fig:12co_channels} but for the \thCOfull{} molecular line high-resolution cube imaged at a channel spacing of 200\,\msec{}.}
    \label{fig:13co_channels}
\end{figure*} 
%%%%%%%%%%%%%%%%%%%%%%%%%%%%%%%%%%%%%%%%

%%%%%%%%%%%%%%%%%%%%%%%%%%%%%%%%%%%%%%%%
\begin{figure*}[t]
    \centering
    \includegraphics[width=1.0\linewidth]{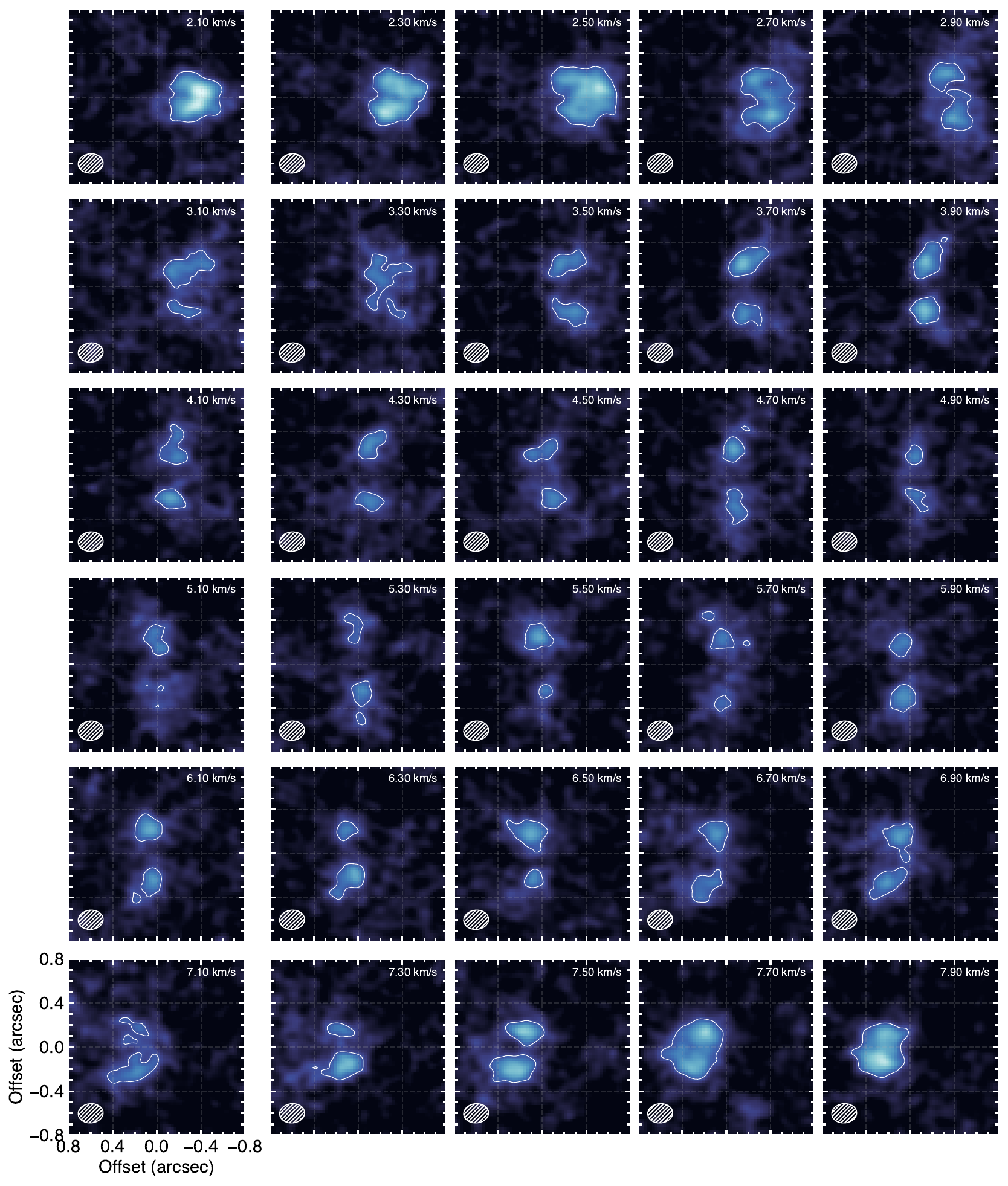}
    \caption{Same as Fig.\,\ref{fig:12co_channels} but for the \eiCOfull{} molecular line with a beam of 0.22\arcsec{}\,x\,0.17\arcsec{} and imaged at a channel spacing of 200\,\msec{}.}
    \label{fig:c18o_channels}
\end{figure*} 
%%%%%%%%%%%%%%%%%%%%%%%%%%%%%%%%%%%%%%%%

%%%%%%%%%%%%%%%%%%%%%%%%%%%%%%%%%%%%%%%%
\begin{figure*}[t]
    \centering
    \includegraphics[width=1.0\linewidth]{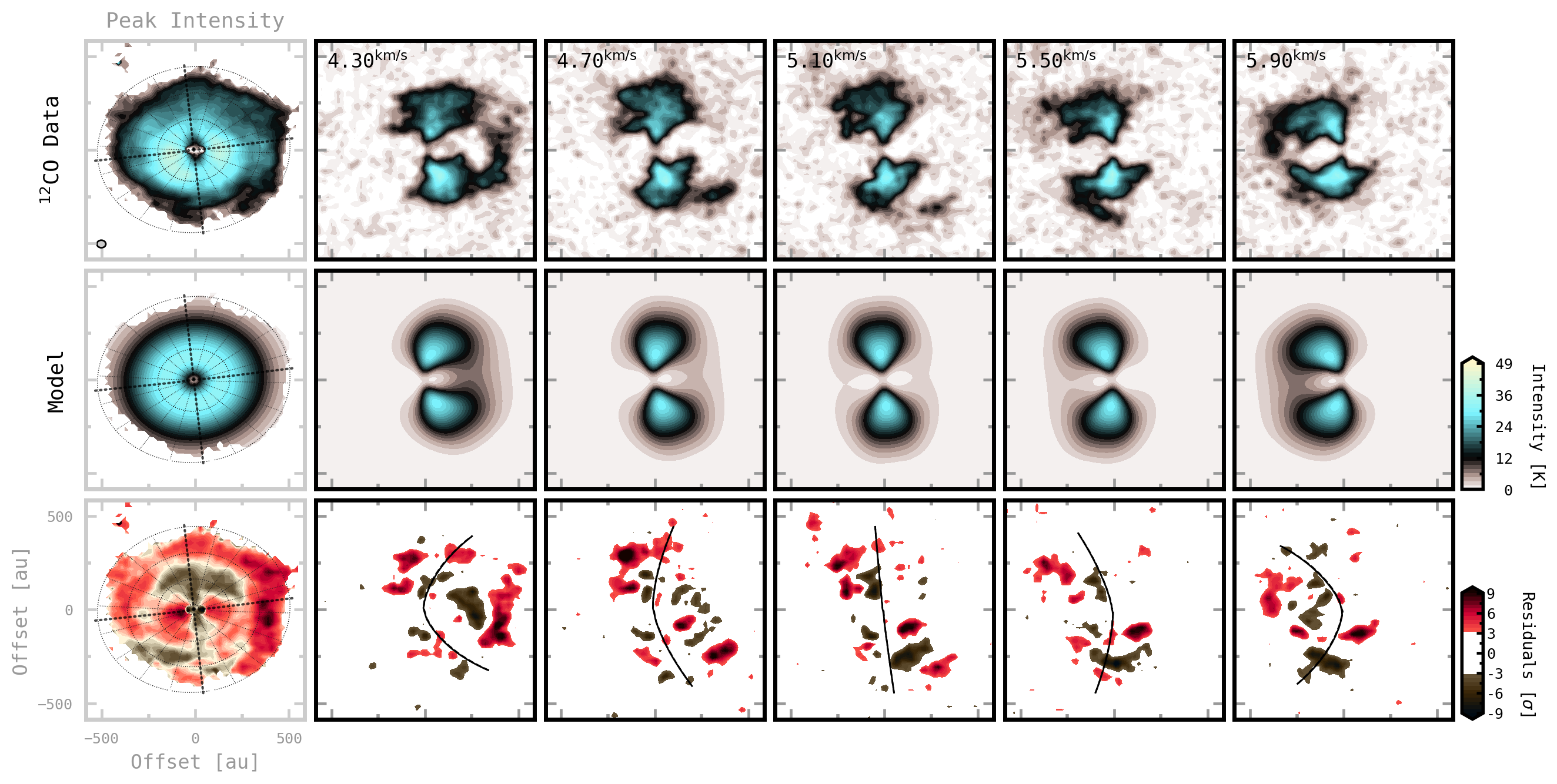}
    \caption{Selected line intensity channels for \twCOfull{} using the high-resolution cubes. The panels for each figure display the data channels, the best-fit model, and the residuals, from top to bottom. Residuals below 3$\sigma=5.0\,$K are whited out. The systemic-velocity channel is at 5.1\kmsec{}, and the left-most panels show the corresponding peak intensity maps.}
    \label{fig:discminer_channels_12co}
\end{figure*} 
%%%%%%%%%%%%%%%%%%%%%%%%%%%%%%%%%%%%%%%%

%%%%%%%%%%%%%%%%%%%%%%%%%%%%%%%%%%%%%%%%
\begin{figure*}[t]
    \centering
    \includegraphics[width=1.0\linewidth]{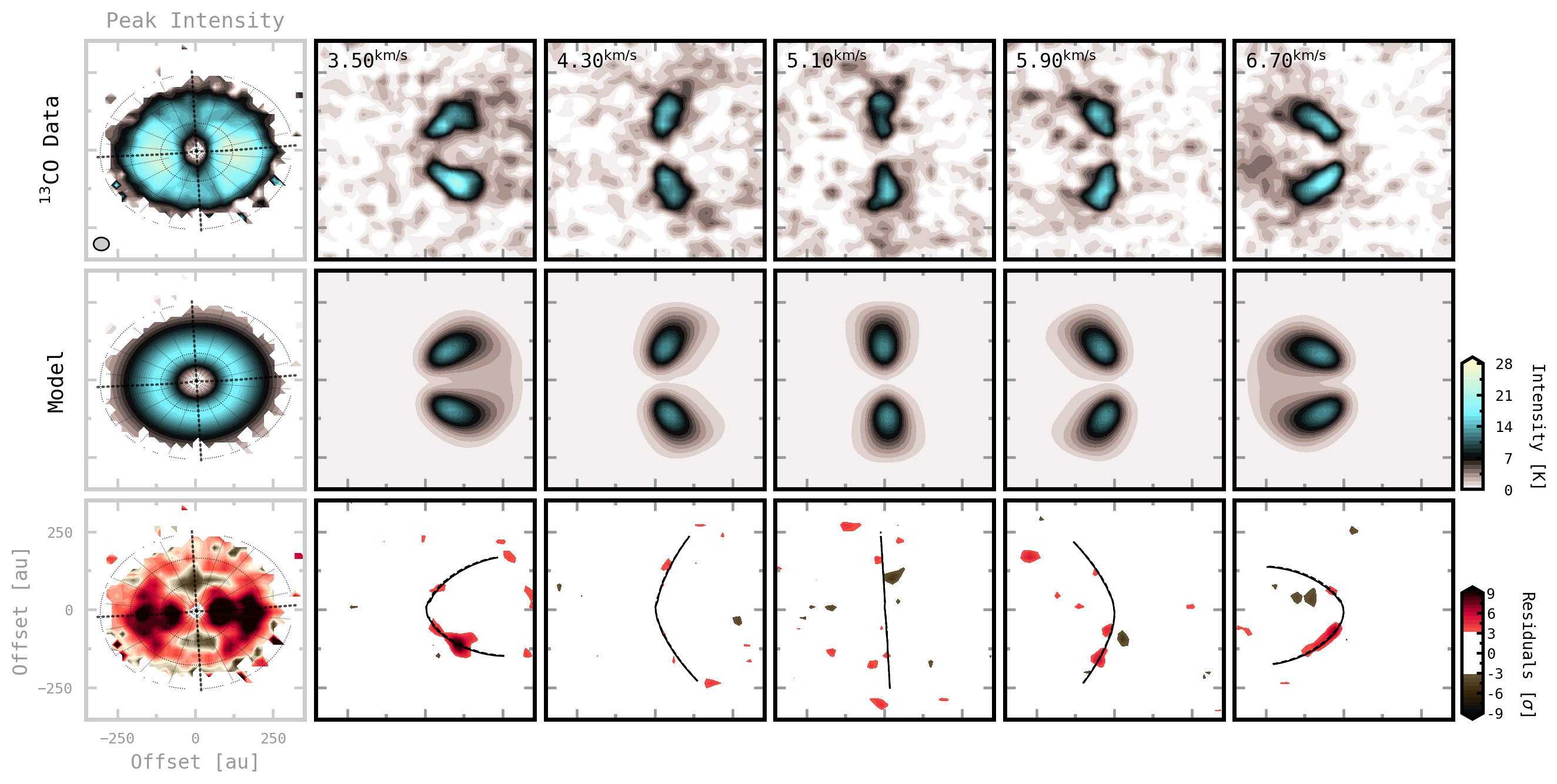}
    \caption{Same as Fig.\,\ref{fig:discminer_channels_12co} but for \thCOfull{} using the high-resolution cube with residuals below 3$\sigma=3.6\,$K are whited out.}
    \label{fig:discminer_channels_13co}
\end{figure*} 
%%%%%%%%%%%%%%%%%%%%%%%%%%%%%%%%%%%%%%%%

%%%%%%%%%%%%%%%%%%%%%%%%%%%%%%%%%%%%%%%%
\begin{figure*}[t]
    \centering
    \includegraphics[width=1.0\linewidth]{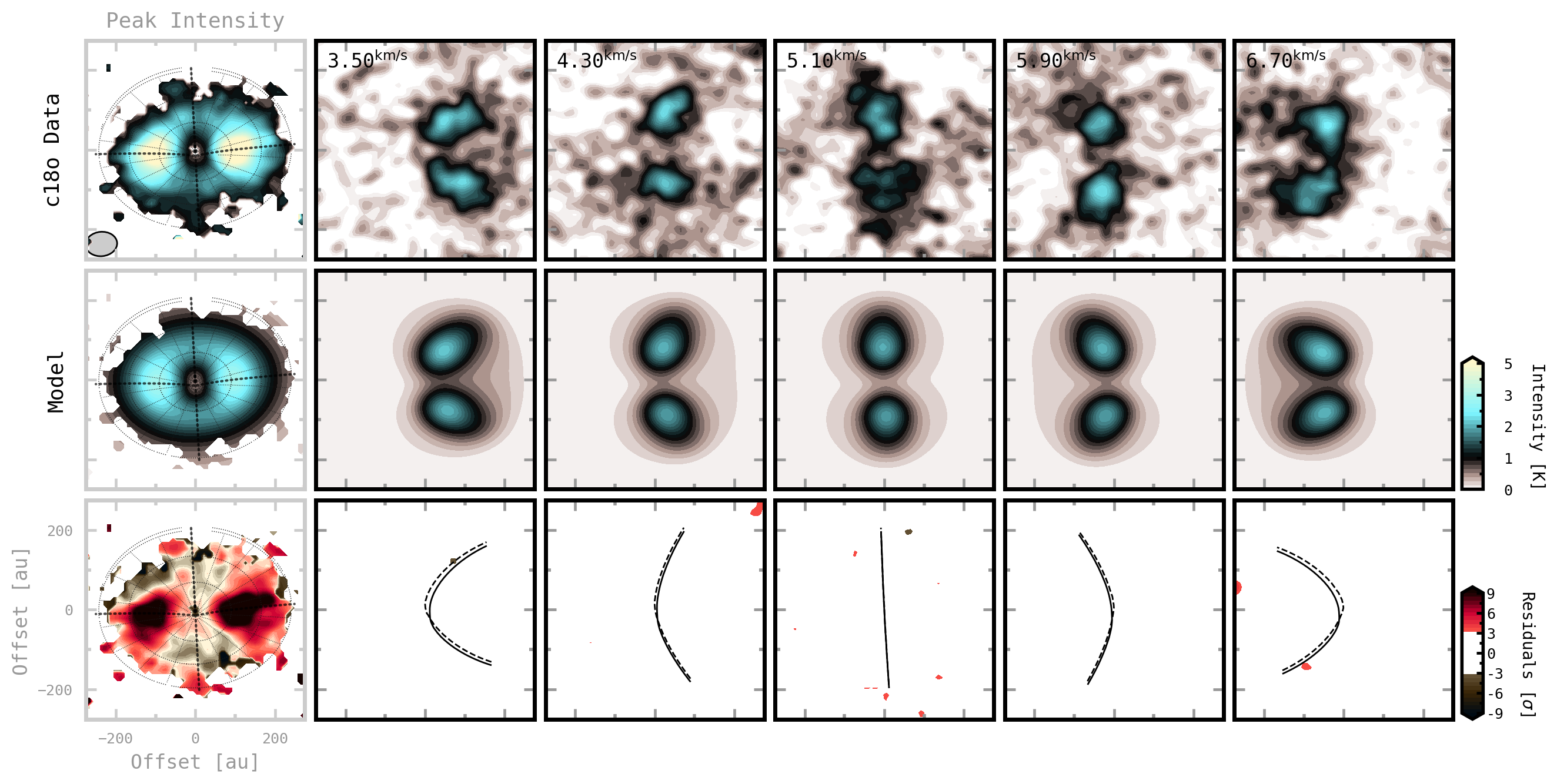}
    \caption{Same as Fig.\,\ref{fig:discminer_channels_12co} but for \eiCOfull{} using the fiducial cube with residuals below 3$\sigma=3.0\,$K are whited out. Note that there are nearly no residuals above $3\sigma$ due to the low S/N of the data.}
    \label{fig:discminer_channels_c18o}
\end{figure*} 
%%%%%%%%%%%%%%%%%%%%%%%%%%%%%%%%%%%%%%%%

%%%%%%%%%%%%%%%%%%%%%%%%%%%%%%%%%%%%%%%%
\begin{figure*}[t]
    \centering
    \includegraphics[width=1.0\linewidth]{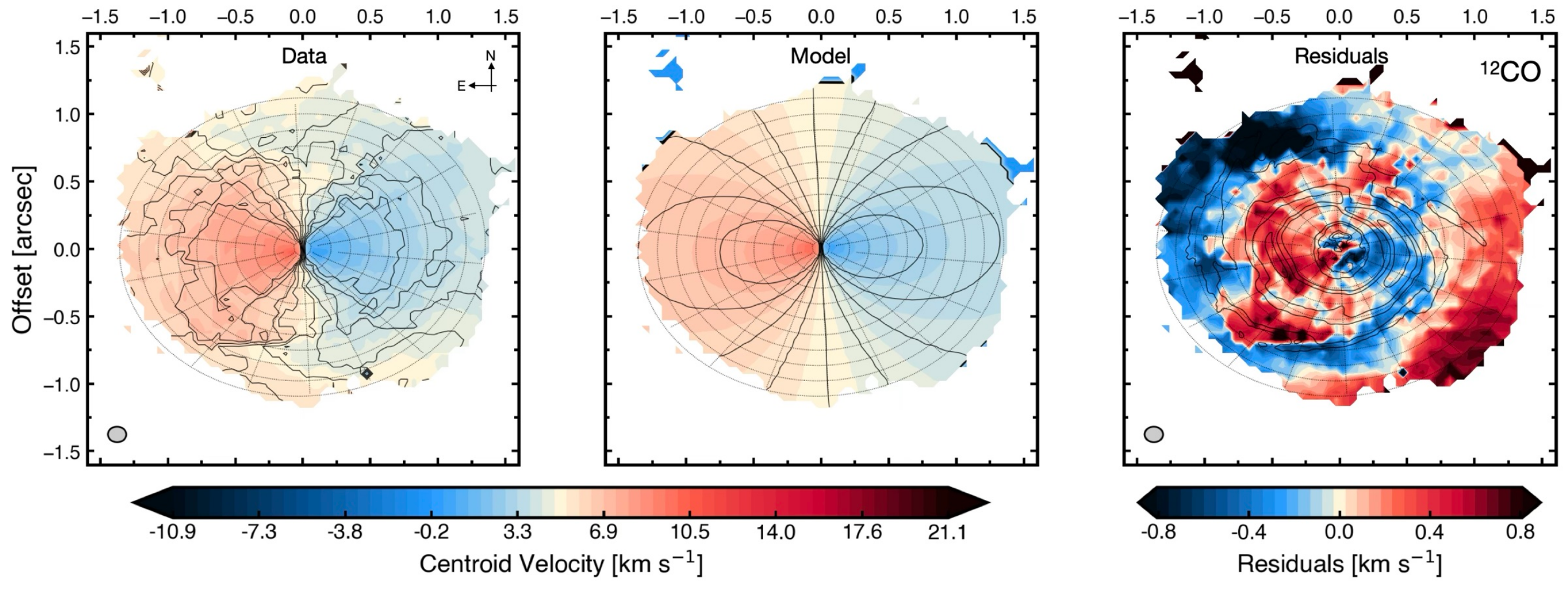}
    \caption{\twCO{} quadratic velocity residuals. Here we use the \texttt{bettermoments} ``quadratic'' centroid moment, which fits a quadratic function to the peak of the line and its two neighboring pixel to determine the line centroid \citep{Teague_2019}. The employed \twCO{} robust\,0.5 cube, Keplerian \texttt{discminer} model, and layout is the same as in Fig.\,\ref{fig:CO_residuals}. The quadratic LoS velocity map of the data differs from the Gaussian one of Fig.\,\ref{fig:CO_residuals}. The ``quadratic'' line centroid traces more closely the peak of the emission, hence leading to strong bending of the iso-velocity contours coinciding with the regions of broadened line-profiles, as highlighted in the map of Fig.\,\ref{fig:line_profiles}. In particular, at the location of the infalling material in the southeast, this leads to a strong blue- or redshifted residual due to the double-peaked line profile at this location.}
    \label{fig:12co_quad_resid}
\end{figure*} 
%%%%%%%%%%%%%%%%%%%%%%%%%%%%%%%%%%%%%%%%

%%%%%%%%%%%%%%%%%%%%%%%%%%%%%%%%%%%%%%%%
\begin{figure*}[t]
    \centering
    \includegraphics[width=0.9\linewidth]{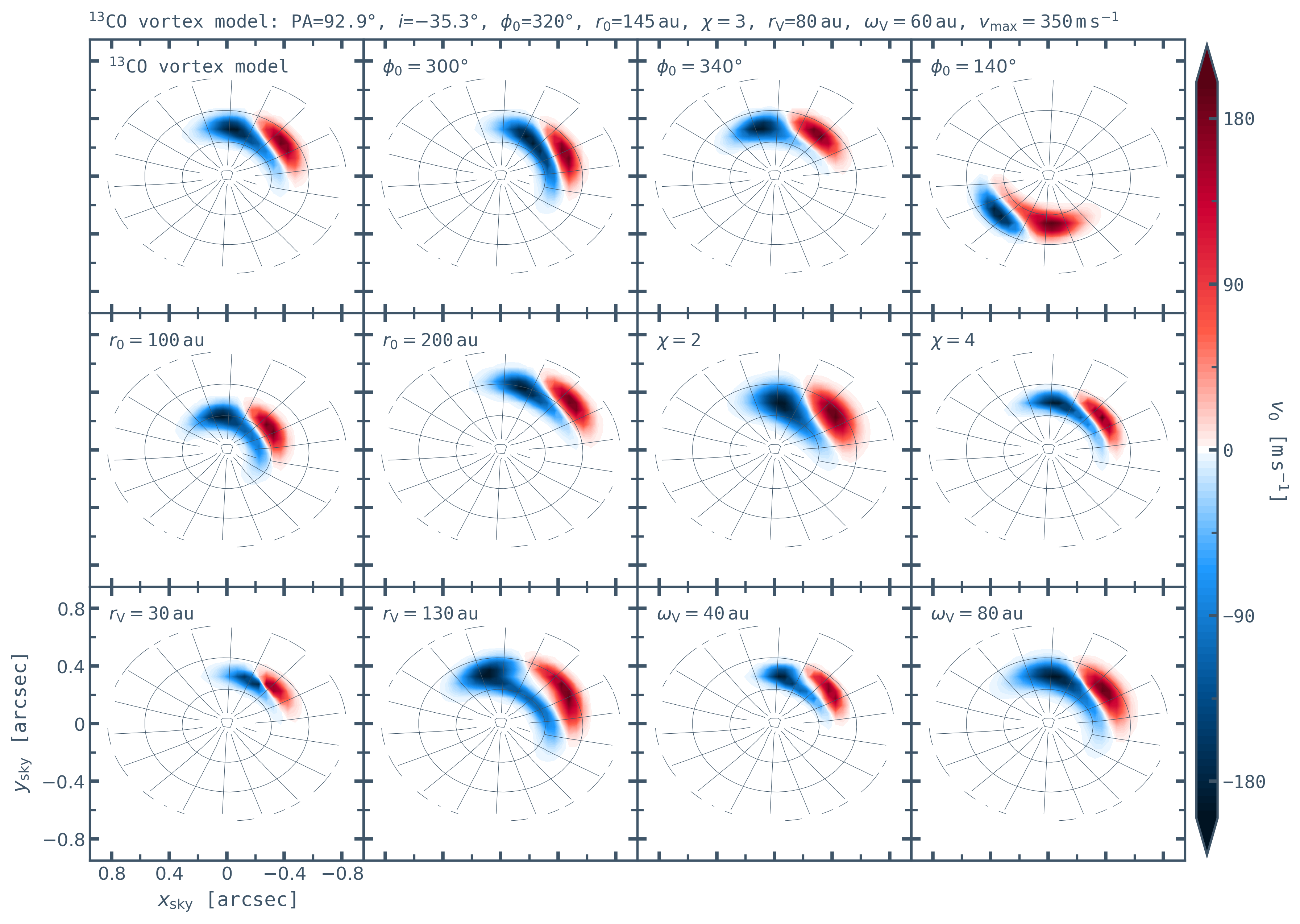}
    \caption{Line-of-sight velocities resulting from the analytical vortex model of \cite{Woelfer_ea_2025}. The model best describing the \thCO{} velocity residuals (Fig.\,\ref{fig:vortex_kinematics}) is shown in the upper left panel with employed parameters listed on the top of the figure. All the other panels vary one of these best-fit parameters to show how the resulting $\upsilon_\mathrm{l.o.s.}$ changes. Position angle and inclination are taken from the \thCO{} best-fit Keplerian model (Table \,\ref{tab:params}). Parameters $\phi_0$ and $r_0$ are the angle and radial center of the continuum crescent. The remaining parameters describe the radius of the ring of maximum velocity ($r_\mathrm{v}$), the Gaussian width of the vortex ring ($\omega_\mathrm{v}$), its aspect ratio ($\chi$, for $\chi=1$ the vortex is circular), and the maximum velocity of vortex rotation ($\rm{v}_\mathrm{max}$; see Fig.\,3 of \citealt{Woelfer_ea_2025} for a sketch).}
    \label{fig:vortex_models}
\end{figure*} 
%%%%%%%%%%%%%%%%%%%%%%%%%%%%%%%%%%%%%%%%

\end{appendix}

\end{document}